%% file: ms.tex
\documentclass[letterpaper,10pt,journal,final]{IEEEtran}
\usepackage[utf8]{inputenc}
\usepackage[T1]{fontenc}
\usepackage[scaled=0.95]{inconsolata}

\usepackage{latexsym,amsmath,amssymb,amsfonts,mathrsfs}
\usepackage{arydshln}   % To support partitioned matrix with dashed lines
\usepackage{mathtools}
\usepackage{amsthm}
\usepackage{empheq}     % to deal with parentheses + subfunctions

\usepackage{algorithm}
\usepackage{algpseudocode}
\makeatletter
\renewcommand{\ALG@beginalgorithmic}{\small}
\makeatother

\usepackage{float}
\usepackage{graphicx}
\usepackage{caption}
\usepackage{subcaption}

\usepackage{comment}
\usepackage{enumitem}  % [label=\roman*]
\usepackage[usenames,dvipsnames]{xcolor}
\usepackage{cite}
\usepackage[colorlinks=true,linkcolor=magenta,citecolor=blue,urlcolor=cyan,
            filecolor=red]{hyperref}
\newtheorem{theorem}{Theorem}
\newtheorem{lemma}[theorem]{Lemma}
\newtheorem{proposition}[theorem]{Proposition}

\theoremstyle{definition}
\newtheorem{definition}[theorem]{Definition}

\newtheorem*{criteria}{Test Criterion}
\theoremstyle{remark}
\newtheorem{remark}{Remark}

\input{userdef-mathsym.tex}

\title{\LARGE \bf
  System Aliasing in Dynamic Network Reconstruction: \\Issues on Low Sampling Frequencies
}
\author{Zuogong Yue$^{1}$, Johan Thunberg$^{1*}$, Lennart Ljung$^{2}$, Ye Yuan$^{3}$ and Jorge Gon\c{c}alves$^{1}$%  <-this % stops a space
  \thanks{This work was supported by Fonds National de la Recherche Luxembourg (Ref.~9247977).}%
  \thanks{$^{1}$ Zuogong Yue, Johan Thunberg and Jorge Gon\c{c}alves are with
    Luxembourg Centre for Systems Biomedicine,
    6, avenue du Swing, L-4367 Belvaux, Luxembourg.}%
  \thanks{$^{2}$ Lennart Ljung is with Department of Electrical Engineering, Link\"oping University, Link\"oping, SE-58183, Sweden.}%
  \thanks{$^{3}$ Ye Yuan is with the School of Automation, Huazhong University of Science and Technology, Wuhan, 430073, China.}%
  \thanks{\hspace*{0mm}$^{*}$For correspondence, \href{mailto:johan.thunberg@uni.lu}{\tt johan.thunberg@uni.lu}.}%
}

\hypersetup{
  pdftitle={
    System Aliasing in Network Reconstruction: Issues on Low Sampling Frequencies},
  pdfauthor={Zuogong YUE},
  pdfcreator={Emacs version 25.1 + AUCTeX version 11.90}}

\begin{document}

\maketitle
\thispagestyle{empty}
\pagestyle{empty}

% ---------------- ABSTRACT ----------------
\begin{abstract}
  Network reconstruction of dynamical continuous-time (CT) systems is motivated
  by applications in many fields. Due to experimental limitations, especially in
  biology, data could be sampled at low frequencies, leading to significant
  challenges in network inference. We introduce the concept of “system aliasing”
  and characterize the minimal sampling frequency that allows reconstruction of
  CT systems from low sampled data.  A test criterion is also proposed to check
  whether system aliasing is presented.  With no system aliasing, the paper
  provides an algorithm to reconstruct dynamic network from data in the presence
  of noise. In addition, when there is system aliasing we perform studies that
  add additional prior information of the system such as sparsity. This paper
  opens new directions in modelling of network systems where samples have
  significant costs. Such tools are essential to process the available data in
  applications subject to current experimental limitations.
\end{abstract}

% ----------------- MAIN --------------------
\section{Introduction}

Many complex systems can be modeled as networks in applications to reveal and
illustrate interactions between measured variables. A common characteristic of
such networks is sparsity, where each variable in only involved in a few
interactions. Such sparse networks are often presented in systems in nature.  An
example of the latter is the interaction between species such as genes and
proteins in human cells; such interactions can be modeled by stochastic/ordinary
differential equations, e.g. \cite{palsson2011systems}.  Such network models in
biology help to understand, for instance, metabolic pathways, interactions
between DNAs/proteins, and furthermore contribute to pathology of disease
detection on or even clinical treatment to complicate diseases,
e.g. \cite{Bar-Joseph2012}.  Motivated by practical applications,
reconstruction of sparse (Boolean) networks turns to be critical as more
techniques have been available to acquire time-series data.

There has recently been quite some interest in the study of dynamic networks
from different perspectives: network
identifiability\cite{Goncalves2008,Hayden2016a}, network module
identifiability\cite{VandenHof2013}, network inference using discrete-time
approaches\cite{Chiuso2012,Yue2017a}, etc. With regard to network inference, the
factor that distinguishes itself from traditional system
identification\cite{Ljung1998} is the particular request on sparse
structures. To enhance sparsity, there are multiple methods are available:
LASSO\cite{Yuan2006}, iterative reweighted $l_1$/$l_2$
algorithms\cite{Candes2008,Chartrand2008}, Sparse Bayesian
Learning\cite{Tipping2001,Wipf2007}, etc.
% Sparse network reconstruction usually uses the group version of these methods.

It deserves to be emphasized that the discrete-time approach for network
inference is valid only if the sampling frequency is high enough, where the
discrete-time model shares the same network structure as the continuous one that
is the physical process (here we assume the dynamical systems evolve in
continuous time in nature).  To use discrete-time methods, one practical rule to
choose sampling frequencies is taking ten times the bandwidth of the underlying,
in this case assumed to be linear, systems\cite{Ljung1998}.  However, in
biological systems, most time-series data are sampled considerably slower than
this empirical frequency, e.g. ``high time-resolution'' time series in
\cite{He2012}, which usually cannot be solved by increasing sampling rates due to
various constraints in biological experiments.

There have been several studies on the identification of continuous-time
systems, e.g. \cite{Ljung2010,Garnier2003}. However, most methods request a high
sampling frequency to guarantee certain simplifications on theoretical
deductions or numerical calculations. Choosing a fairly low sampling frequency
may trigger the problem of ``system aliasing'', that is, multiple
continuous-time systems produce exactly the same output samples, while having
different network structures. To determine physical interconnections, it is
inevitable to resorting to the identification of continuous-time models.  With decrease of sampling frequencies, it becomes particularly
challenging, nearly intractable, in theory and computation to identify sparse
structures of continuous-time models.

In this paper, we first reveal the challenges due to the low sampling frequency
by examples in Section~\ref{sec:syst-alias-sysid} and then present a definition
of \emph{system aliasing}. A Nyquist-Shannon-like sampling theorem is presented
in Section~\ref{sec:no-system-aliasing} to determine the minimal sampling
frequency that avoids the effect of system aliasing.  Section~\ref{sec:methods}
presents an algorithm to reconstruct sparse networks in the case of no system
aliasing using low-sampling-frequency data. The case with system aliases is
discussed in Section~\ref{sec:syst-alias-bound-constraints}, which discusses the
feasibility of exploring the ground truth in theory. The last section,
Section~\ref{sec:simulations}, provides numerical examples to show performance
of the proposed methods.

\section{Problem Formulation}
\label{sec:problem-formulation}

Consider a filtered probability space
$(\Omega, \mathcal{F}, (\mathcal{F}_t)_{t \in [0, \infty)}, \mathbb{P})$, where the
filtration is always assumed to be complete.  Let $\{w(t): t \geq 0\}$ be the
$n$-dimensional standard $\mathcal{F}_t$-Brownian motion.
The physical plant/process in our study, as a dynamical
system in continuous time, is modeled by the following stochastic differential
equation
\begin{equation}
  \label{eq:dyna-sys-ss-cont}
  \mathrm{d}x = Ax\,\mathrm{d}t + Bu\,\mathrm{d}t + R \mathrm{d}w,
\end{equation}
where $A \in \mathbb{R}^{n \times n}$ is stable, $R \in \mathbb{R}^{n \times n}$ is
symmetric and positive definitive, the initial $x(t_0)$ is a
Gaussian random variable with mean $m_0$ and variance $R_0$, $t_0 \geq 0$, and $w(t)$ is interpreted
as \emph{disturbance} on the state variables (or called \emph{process noise}).  The solution
to \eqref{eq:dyna-sys-ss-cont} is an $\mathcal{F}_t$-adapted $n$-dimensional
stochastic process $x(t) = (x_1(t), \cdots, x_n(t))_{t \geq t_0}$ such that
\begin{equation*}
  x(t) = x(t_0) + \int_{t_0}^t \left( A x(s) + B u(s)  \right)\, \mathrm{d}s + \int_{t_0}^t R \mathrm{d} w(s),
\end{equation*}
where $x(t_0) \in \mathcal{F}_{t_0}$ and see \cite{Gall2016} for the definition of
stochastic integral. It is assumed that $x(t_0), w(t)$ are independent. The solution
$x(t)$ is \emph{strong}, that is, $x(t)$ is adapted to
$\mathcal{F}_t^w \coloneqq \sigma(w_u\! :\! u\!\leq\! t) \vee \sigma(\mathcal{N})$
(i.e. the complete $\sigma$-field generated by $\{w_u\!:\! u\!\leq\!  t\}$; see
\cite{Gall2016} for details).  An input signal $\{u(t), t \geq t_0\}$ has been
applied to the system, and the output $y$ of the system is observed at the discrete
times $t_0, t_1, ..., t_N$,
\begin{equation}
  \label{eq:dyna-sys-ss-output}
  y(t_k) = C x(t_k)
\end{equation}
where $C = [I\; 0] \in \mathbb{R}^{p \times p},\ p \!\leq\! n$,
% the \emph{measurement noise} $\{\varepsilon(t_k)\}$ is Gaussian i.i.d. with zero mean and variance $R_2$,
$t_k \triangleq t_0 + kh$ and $h > 0$ is the sampling period. Here the measurement
noise is not included mainly due to that we have not yet given a definition of
network models (see \cite{Goncalves2008}) from state-space representations with
measurement noises.  The stochastic difference equation that relates the values of
the state variable $x$ in \eqref{eq:dyna-sys-ss-cont} at the sampling
instants\cite[p.~82-85]{Astrom2012}\cite[chap.~2]{Garnier2008} is given by
\begin{equation}
  \label{eq:dyna-sys-ss-discr}
    x(t_{k+1}) = A_dx(t_k) + B_d u(t_k) + v(t_k),
\end{equation}
where
\begin{equation}
  \label{eq:formula-Ad-Bd}
  A_d = \exp({hA}), \quad
  B_d = \int_0^h \exp(sA)B \,\mathrm{d}s,
\end{equation}
$\exp(\cdot)$ is the matrix exponential, and the Gaussian i.i.d. $v(t)$ has mean zero
and covariance matrix
\begin{equation}
  \label{eq:disc-time-sys-wn-cov}
  R_{d} = \int_{0}^h \exp({sA}) R \exp({sA^T}) \,\mathrm{d}s.
\end{equation}
% In the whole paper, if not specified explicitly, we reserve $A$ for the ground truth and $A_d \triangleq \exp(hA)$.

The \emph{linear dynamic network} model of \eqref{eq:dyna-sys-ss-cont} is given as
\begin{equation}
  \label{eq:dsf-cont}
  y(t) = Q(q) y(t) + P(q) u(t) + H(q) e(t),
\end{equation}
where $Q(q),P(q)$ and $H(q)$ are $p \times p$, $p\times m$ and $p \times p$ matrices of
strictly-proper real-rational transfer functions respectively in terms of $q$, $q$ is
the differential operator $q x(t) = \mathrm{d}x/\mathrm{d} t$, and $e(t)$ is the
Gaussian white noise with zero mean and $\mathbb{E}[e(t)^T e(s)] = I\delta(t-s)$
(e.g. see \cite{Hayden2016a}). The model \eqref{eq:dsf-cont} is called
\emph{Dynamical Structure Function} (DSF), firstly proposed in \cite{Goncalves2008}.
The network model defines path diagrams which show the interconnections between the
elements of the output variable.
\begin{definition}[\cite{Yue2017a}]
  \label{def:LTI-dynamic-network}
  Let $\mathcal{G} = (V,E)$ be a digraph,
  where the vertex set $V = \{y_1, \dots, y_p, u_1, \dots, u_m\}$
  % \footnote{Here $y_i, w_k$ are label names of the vertices, instead of signals $y_i(t), w_k(t)$.},
  and the arc (directed edge) set $E$ is defined by
  \begin{enumerate}[label=\roman*),itemindent=8pt]
  \item $(y_j, y_i) \in E\;\, \Leftrightarrow\; Q_{ij}(q) \neq 0$,
  \item $(u_k, y_i) \in E\; \Leftrightarrow\; P_{ik}(q) \neq 0$,
  \item $(y_i, u_k) \notin E, \; \forall i,k$.
  \end{enumerate}
  Let $f$ be a map defined as
  \begin{equation*}
    \begin{array}{rrll}
      f: & E & \rightarrow & S_\text{TF} \\
         & (y_j, y_i) & \mapsto & Q_{ij}(q) \quad \mathrm{or} \quad (u_k, y_i)  \mapsto  P_{ik}(q),
    \end{array}
  \end{equation*}
  where $S_\text{TF}$ is a subset of single-input-single-output (SISO) (strictly) proper real rational transfer functions.
  We call the tuple $\mathcal{N} \coloneqq (\mathcal{G}, f)$ a (linear) \emph{dynamic network},  $f$ the (linear) \emph{dynamic capacity function} of $\mathcal{N}$, and $\mathcal{G}$ the \emph{underlying digraph} of $\mathcal{N}$, which is also called (linear) \emph{Boolean dynamic network}.
\end{definition}

This article focuses on the full-state measurement case, i.e. $C = I$, where $A$ in
\eqref{eq:dyna-sys-ss-cont} coincides with $Q(q)$ in network models
\eqref{eq:dsf-cont}.
% In full-state measurement cases, it may benefit little from modeling noise
% separately in terms of process noise and measurement noise, and the network model
% does not distinguish types of noise. However, including measurement noise brings
% particular challenges in network reconstruction from low sampling frequency data
% (see Remark~\ref{rmk:challenge-from-measurement-noise}). Hence, the model in
% Section~\ref{sec:methods} only includes process noise, which might also be
% considered as a lump-sum of noise effects.
Concerning the network identifiability\cite{Goncalves2008}, we assume $B$ to be
diagonal\footnote{This is not required in system identification, in which only the
  input-output behavior is concerned. Any state-space realization could be feasible
  solutions.} (DSF, \cite{Goncalves2008}) or particularly $B = I$ (the model used in
\cite{VandenHof2013}).  Let the measurement be denoted by
\begin{math}
  Y^N \triangleq [ y(t_0), \; y(t_1), \; \dots, \; y(t_N) ].
\end{math}
We summarize the main problem in our study as follows:

\smallskip
\noindent\textbf{Main Problem}:
Given the finite signal $Y^N$ in full-state measurement (i.e. $C=I$), with probably
large $h$ (the sampling period) and small $N$ (the length of time series), infer the
dynamic network $\mathcal{N}$ (or the Boolean $\mathcal{G}$), assuming that the
ground truth $A$ is sparse and and $B$ is diagonal.
\medskip

\begin{remark}
  \label{rmk:prob-hard-two-reasons}
  The problem is challenging due to the following two major reasons:
  \begin{itemize}
  \item Since $h$ could be large, i.e. the sampling frequency is low, we have to
    estimate $A$ in order to determine $\mathcal{G}$ or $\mathcal{N}$ (see
    Figure~\ref{fig:net-topol-A-Ad}).
  \item Since $N$ does not approach infinity, the estimation of $A_d$ from PEM
    (Prediction Error Minimization) or ML (Maximum Likelihood) may fail to
    identify $\mathcal{G}$ correctly by taking matrix logarithms, even though
    PEM/ML gives consistent estimation in theory.
    % One has to notice that, even if $\hat{A}_d$ is close to $A_d$ in the sense of
    % prediction, $\mathcal{G}$ determined from the matrix logarithm of $\hat{A}_d$ may
    % significantly different from the ground truth.
    % (i.e. $\lim_{N \uparrow +\infty}\mathbb{E}(\hat{A}_d(X^N)) = A_d$)
  \end{itemize}
  % Moreover, note that $\{v(t): t = 1, 2, \dots\}$ is not the ``random noise''
  % excitation signal that is sampled in many system identification experiments. It is
  % called \emph{blind identification} to emphasize the difference
  % (e.g. \cite{Hua2000}).
\end{remark}

Throughout the text, by default, we always deal with \emph{primary} matrix functions,
including $\exp$ (matrix exponential), $\log$ (matrix logarithm) and $\Log$ (the
principal matrix logarithm in Theorem~\ref{thm:log-uniq-matrix}). \emph{Primary}
matrix functions refer to the ones defined via \emph{Jordan Canonical Form} or
equivalently via \emph{Polynomial Interpolation}, \emph{Cauchy Integral Theorem}
\cite[chap.~1]{Higham2008}. The \emph{primary} notion of matrix functions is of
particular interest and the most useful in applications\cite{Higham2008,Horn2003}.

\section{System Aliasing in Identification}
\label{sec:syst-alias-sysid}

\subsection{Observations on matrix logarithm}
\label{subsec:ambig-matrix-log}

Supposing that ${A}_d$ has been perfectly estimated from samples, the estimate of the
$A$ matrix for the continuous-time system is straightforwardly calculated by solving
\begin{equation}
  \label{eq:exp-hA-eq-Ad}
  \exp(h {A}) ={A}_d.
\end{equation}
via matrix logarithm.  However, referring to Theorem~\ref{thm:matr-logar-Gantmacher}
\cite{Higham2008}, the equation \eqref{eq:exp-hA-eq-Ad} has several (in fact infititely many)
solutions. Let us review the following observations on \eqref{eq:exp-hA-eq-Ad} to see
the troubles from low sampling frequencies (i.e. $1/h$).

\begin{figure}[htb]   %htbp!H
  \centering
  \includegraphics[width=.3\textwidth]{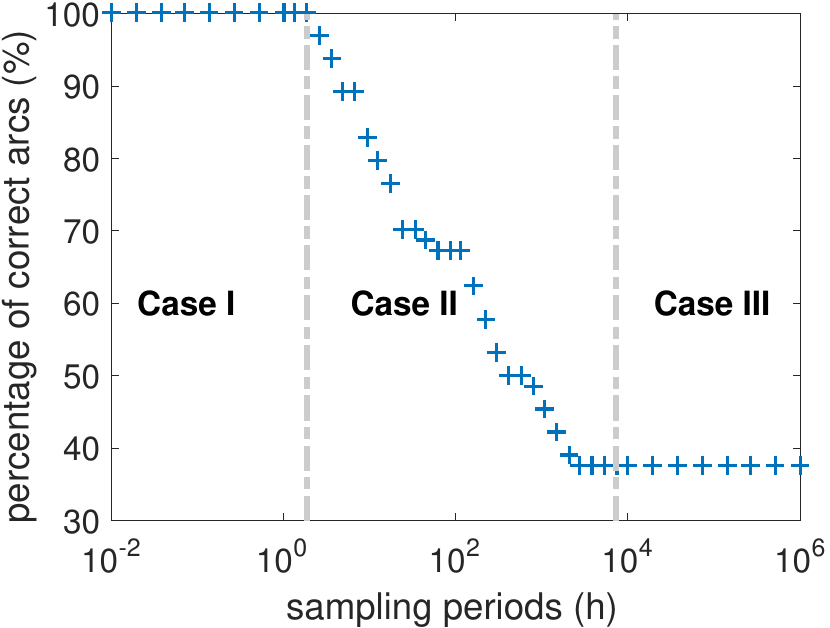}
  \caption{This example randomly chooses a sparse $A$ and $A_d = \exp(hA)$. The
    Boolean networks are determined from $A$ and $A_d-I$ by
    Definition~\ref{def:LTI-dynamic-network} (treating $A$ and $A_d-I$ as $Q$'s),
    denoted by $\mathcal{G}(A), \mathcal{G}(A_d-I)$. We use $\mathcal{G}(A)$ as the
    ground truth. Then $\mathcal{G}(A_d -I)$ is compared with $\mathcal{G}(A)$ and
    the same arcs are labeled as the correct ones.}
  \label{fig:net-topol-A-Ad}
\end{figure}

\smallskip
\noindent\underline{Observation 1}: With the increase of $h$, the Boolean structures of
$A$ (i.e. $\mathcal{G}$ determined from $A$) and $A_d - I$ become more and more different, as illustrated by Figure~\ref{fig:net-topol-A-Ad}.
  The sampling frequency ($1/h$) deserves to be emphasized as a core factor in the
  categorization of different cases in our study:
  \begin{itemize}
  \item {Case~I}: when $h$ is ``very small'' such that $A_d$ shares the same Boolean
    structure as $A+I$. Indeed, one can see it by
    $\exp(hA) = I + hA + \frac{h^2}{2!} A^2 + \cdots$. Hence we can determine
    $\mathcal{G}$ by identifying discrete-time models;
  \item {Case~II}: when $h$ is ``large'' but the ground truth $A$ is
    still the principle matrix logarithm of $A_d$;
  \item {Case~III}: when $h$ is ``even larger'' such that the ground truth $A$ is no
    longer the principle logarithm of $A_d$.
  \end{itemize}
  % The strict quantification of $h$ for the above categorization is presented in
  The general network model \eqref{eq:dsf-cont} of {Case~I} has been solved by
  discrete-time approaches , e.g. see \cite{Yue2017a,Chiuso2012}.  {Case~II}
  is what we mainly studied in this paper. We call both {Case~I} and
  {II} \emph{no system aliasing}, as defined and studied in later sections.
  % Section~\ref{sec:syst-alias-sysid} and \ref{sec:no-system-aliasing}.
  % With regard to {Case~III} which is extremely challenging, we only present
  % certain theoretical results instead of any robust reconstruction
  % approaches. The
  % challenges in {Case~II} and {III} will be further illustrated by
  % examples in Section~\ref{subsec:ambig-matrix-log}.

\begin{figure}[htb]
  \centering
  \includegraphics[width=.42\textwidth]{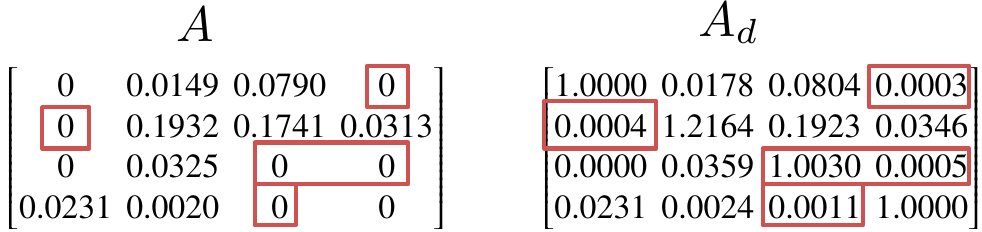}
    \caption{$h = 1$, $A_d = \exp{(hA)}$, $A = \Log{A_d}/h$. }
    \label{fig:logm-eg-1}
\end{figure}

\smallskip
\noindent\underline{Observation 2}: Provided with a sparse $A$, the corresponding $A_d$
could be dense, as the example in Figure~\ref{fig:logm-eg-1}.

\begin{figure}[htb]
  \centering
    \includegraphics[width=.42\textwidth]{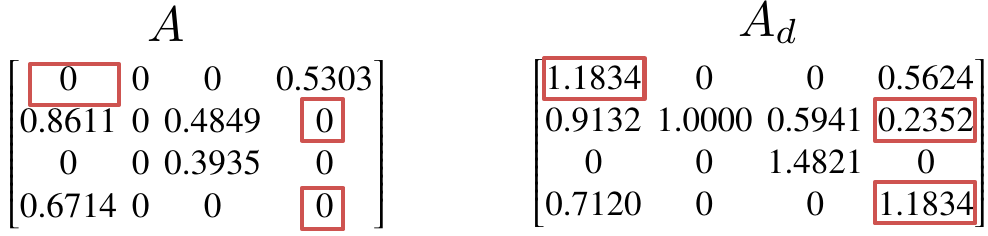}
    \caption{$h = 1$, $A_d = \exp{(hA)}$, $A = \Log{A_d}/h$. }
    \label{fig:logm-eg-2}
\end{figure}

\smallskip
\noindent\underline{Observation 3}: Provided with a sparse $A$, the corresponding $A_d$
can be also sparse. However, they have different Boolean structures (i.e. zeros at
different positions), as shown in Figure~\ref{fig:logm-eg-2}.

\begin{figure}[htb]
  \centering
    \includegraphics[width=0.48\textwidth]{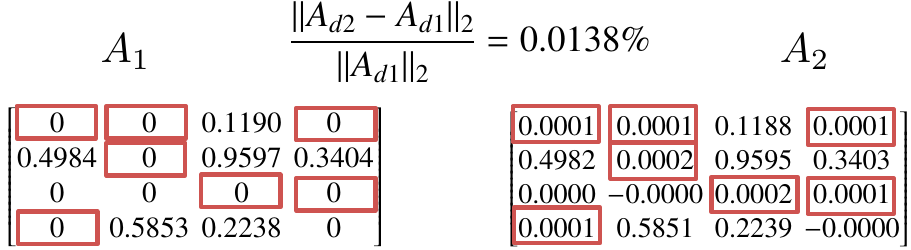}
    \caption{$h = 1$, $A_1 = \Log{A_{d1}}/h$ and $A_2 = \Log{A_{d2}}/h$.}
    \label{fig:logm-eg-4}
\end{figure}

\smallskip
\noindent\underline{Observation 4}: Even though the norm difference of $A_{d1}, A_{d2}$ has been
very small, their matrix logarithms, e.g. $A_1$ and $A_2$ in
Figure~\ref{fig:logm-eg-4}, have significantly different Boolean structures.

\begin{figure}[htb]
  \centering
    \includegraphics[width=0.48\textwidth]{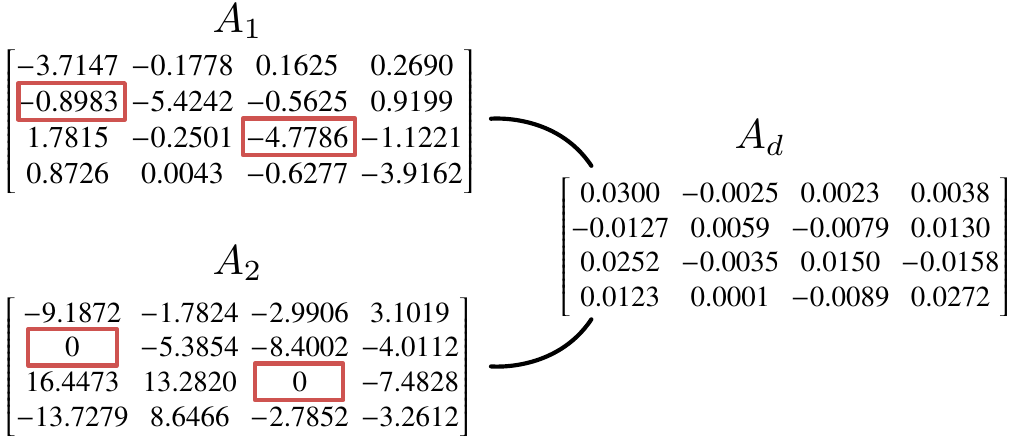}
    \caption{$h = 1$, $\exp{(hA_1)} = \exp{(hA_2)}$, $A_1 = \Log{A_d}/h$, and
      $A_2$ is the ground truth $A$. }
    \label{fig:logm-eg-3}
\end{figure}

\smallskip
\noindent\underline{Observation 5}: There exists more than one solution, e.g. $A_1$ and
$A_2$ in Figure~\ref{fig:logm-eg-3}, and they have different Boolean structures.

Considering the problem formulation in Section~\ref{sec:problem-formulation}, one may
have already noticed the troubles on network reconstruction, which originate from the
matrix logarithms, due to the low sampling frequency.  The examples in
{Observation~1} clearly show that why we have to resort to the continuous-time
system identification to infer network structures. {Observation~2} and
{3} tell that there is no consistent relation between the sparsity of $A$ and
$A_d$.  {Observation~4} points out that the Boolean structures of the
principle logarithms of two $A_{d1}$ and $A_{d2}$ close in matrix norms could be
significantly different.  The example on {Observation~5} shows an even worse
case: the sample period is so large that the principle matrix logarithm is no longer
$A$, which appears as other branches of matrix logarithm of $A_d$, in which no robust
algorithm has yet been available.

\begin{remark}
  In a sum of the above observations, it tells us that, in the identification of $\mathcal{G}$ or $\mathcal{N}$,
  \begin{itemize}
  \item  $\mathcal{G}$ should be determined from $A$ instead of $A_d$, when $h$ is
    large (w.r.t. $A$); (see {Observation~1,~4})
  \item  the sparsity penalty has to be imposed on $A$ directly instead of $A_d$,
    when $h$ is large (w.r.t. $A$); (see {Observation~2,~3,~4})
  \item  the $A$ matrix should be estimated directly instead of via taking matrix
    logarithm of $A_d$, in the presence of noise and a limited length of
    signals. (see {Observation~4})
  \end{itemize}
\end{remark}

\subsection{Definitions}
\label{subsec:definitions}

As shown in Section~\ref{subsec:ambig-matrix-log}, the $A$-matrix has to be
identified in network reconstruction when the sampling frequency is low. In this
scenario, a ``good'' case is that the ground truth $A$ stays as the principle
matrix logarithm of $A_d$ (i.e. {Case~II}); otherwise, it becomes particularly
challenging (i.e. {Case~III}), e.g. Figure~\ref{fig:logm-eg-3}.  To clarify this
classification, we consequently present an important concept in network
reconstruction with low sampling frequencies, ``\emph{system aliasing}''.

Let $\kvec(X)$ denote the vectorization of the matrix $X$ formed by stacking the
columns of $X$ into a single column vector; and $\ikvec(\cdot)$ is defined by
$\ikvec(\kvec(X)) = X$. $\im(x)$ denotes the imaginary part of the complex
number or vector $x$.
\begin{definition}
  \label{def:E-for-sys-alias}

  \begin{align*}
    &\mathscr{E}(A, M, h, \mathscr{S})=
      \Big\{A^* \in \mathbb{R}^{n \times n}:\ M \in \mathbb{R}^{n^2 \times n^2}, h \in \mathbb{R}, \\
    \nonumber
    & A^*  =  \mathrm{arg}\min_{\tilde{A} \in \mathscr{S} }\|M\,\mathrm{vec}(\mathrm{exp}(hA))
      - M\,\mathrm{vec}(\mathrm{exp}(h\tilde{A}))\|_{2} \Big\},
  \end{align*}
  where $\mathscr{S} \subseteq \mathbb{R}^{n \times n}$ contains $A$.
\end{definition}

With this general notation, we present a definition of \emph{system aliasing}
only in terms of the $A$ matrix in state-space representations and the sampling
period $h$, which does not depend on specific identification methods or data.
Before presenting the concept of system aliasing, we have to assume no loss of
information of input signals during sampling, e.g. no inputs, or the continuous
input signal can be determined by input samples together with, for instance, the
zero-order holder. Otherwise, we have to include constraints of input signals in
our definition, which has not yet been studied.

\begin{definition}[System aliasing]
  \label{def:system-aliasing}

  Given $A \in \mathscr{S}$ and $h \in \mathbb{R}_+$, if there exists
  $\hat{A} \neq A \in \mathscr{E}(A, I, h, \mathscr{S})$ and $\hat{A}$ is called
  \emph{system alias} of $A$ with respect to $\mathscr{S}$.  By default, we
  choose
  $\mathscr{S} = \mathscr{S}_A \coloneqq \big\{ \tilde{A} \in \mathbb{R}^{n
    \times n}: \max\{\im(\eig(\tilde{A}))\} \leq \max\{\im(\eig({A}))\} \big\}$.
\end{definition}

We are particularly interested in $\mathscr{E}(A, I, h, \mathscr{S}) = \{A\}$,
i.e. there is no issue of \emph{system aliasing}.  Note that the concept of
\emph{system aliasing} does not depend on specific data. It only depends on
system dynamics (e.g. the $A$-matrix in \eqref{eq:dyna-sys-ss-cont}) and
sampling frequencies.  If the $M$ matrix is specifically constructed by data
instead of $I$, $\mathscr{E}(A, M, h, \mathscr{S}) = \{A\}$, where $A$ denotes
the ground truth, tells that the underlying system is identifiable from the
given data (see \cite[Sec.~III-B]{Yue2016b}).  Obviously if we have {system
  aliasing} for the system with a specific sampling frequency, without extra
prior information on $A$, the system is always not identifiable.

\section{No system aliasing: the minimal sampling frequency}
\label{sec:no-system-aliasing}

% \subsection{The minimal sampling frequency}
% \label{subsec:sampl-freq-no-sys-alias}

Provided with the definition of \emph{system aliasing}, a question comes first: what $(A, h)$ satisfies $\mathscr{E}(A, I, h, \mathscr{S}_A) = \{A\}$. To answer this question, we need to introduce a theorem on matrix logarithm.
\begin{theorem}[principal logarithm {\protect \cite[Thm.~1.31]{Higham2008}}]
  \label{thm:log-uniq-matrix}
  Let $P \in \mathbb{C}^{n \times n}$ have no eigenvalues on $\mathbb{R}^-$. There is a unique logarithm $A$ of P all of whose eigenvalues lie in the strip $\{ z: -\pi < \im(z) < \pi \}$. We refer to $A$ as the principal logarithm of $P$ of write $A = \Log(P)$. If $P$ is real then its principal logarithm is real.
\end{theorem}

To make the principal matrix logarithm $\Log(\cdot)$ be well-defined, we always assume
that $\exp(hA)$ has no negative real eigenvalues.  Let
$\mathscr{G}(h) = \{z \in \mathbb{C}: -\pi/h < \im(z) < \pi/h, h \in \mathbb{R}\}$.
By Theorem~\ref{thm:log-uniq-matrix} and \ref{thm:matr-logar-Gantmacher}, it always
holds that $\Log(\exp(h A))/h \in \mathscr{E}(A, I, h, \mathscr{S}_A)$. To avoid
\emph{system aliasing}, it implies that $\Log(\exp(h A))/h = A$ , i.e.
$\eig(A) \in \mathscr{G}(h)$. It is summarized as the following lemma.

\begin{lemma}
  \label{lemma:A=Ahat}
  Let $A_d = \exp(hA), h\in\mathbb{R}_+$, which has no negative real eigenvalues,
  and $\hat{A} = \Log(A_d)/h$. Then $A = \hat{A}$
  (i.e. $\mathscr{E}(A,I,h,\mathscr{S}_A) = \{A\}$) if and only if
  $\eig(A) \in \mathcal{G}(h)$.
\end{lemma}

\begin{figure}[htb]
 \centering
 \includegraphics[width=.4\textwidth]{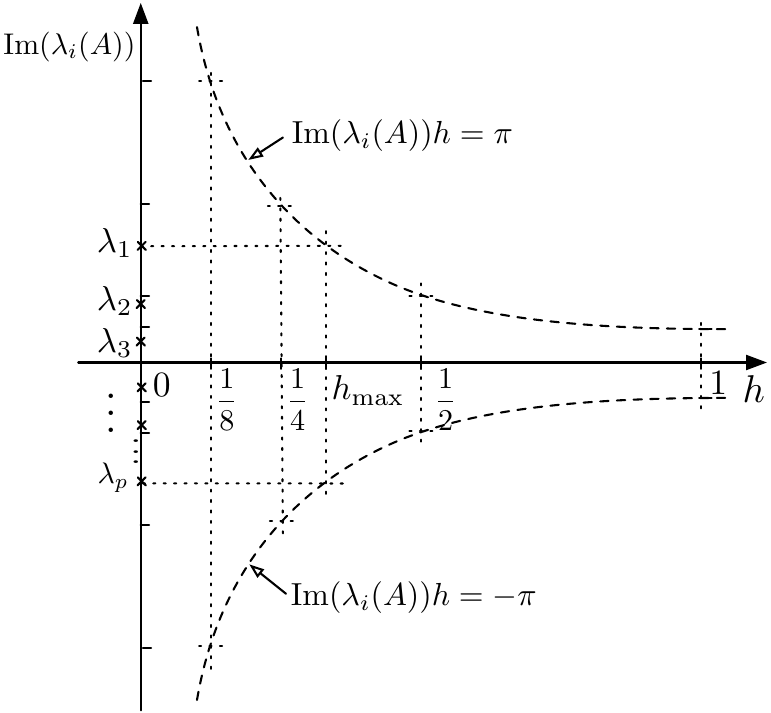}
 \caption{The imaginary parts of all eigenvalues of $A$ must lie into $(-\pi/h, \pi/h)$. $\lambda_i(\cdot)$ denotes the $i$-th eigenvalue of $A$ in Theorem~\ref{thm:matr-logar-Gantmacher}. The symbols ``{\tiny$\times$}'' denote the locations of $\im(\lambda_i(A))$. $h_\mathrm{max}$ is the maximal sampling period that allows taking principal logarithms to estimate $A$, without facing troubles from system aliasing.}
 \label{fig:h-vs-eigv-identifi}
\end{figure}

Given no other information on the system, consider the identification problem of $A$
using full-state measurement. It is necessary to decrease the sampling period $h$
until the ground truth falls into the strip of $\mathscr{G}(h)$, and then the
principal logarithm refers to the ground truth $A$, as illustrated in
Figure~\ref{fig:h-vs-eigv-identifi}. Otherwise, we would be bothered by \emph{system
  aliases} of $A$ and be unable to make a decision, unless we know extra prior
information on $A$.

\begin{theorem}[Nyquist-Shannon-like sampling theorem]
  \label{thm:identifi-largest-sampl-period}
  Considering equidistant sampling, to uniquely reconstruct the continuous-time
  system ${A}$ from the corresponding discrete-time system $A_d$ by taking the
  principal matrix logarithm, the sampling frequency $\omega$ \textup{(rad/s)}
  must satisfy
  \begin{displaymath}
    \omega \geq 2\max \left\{ |\im\left(\lambda_i({A})\right)|,\ i = 1,\dots,n \right\}.
  \end{displaymath}
  Equivalently, the sampling period $h$ (i.e. $2\pi/\omega$) should satisfy
  \begin{displaymath}
    h \leq \min \left\{ \pi / |\im\left(\lambda_i({A})\right)|,\ i = 1,\dots,n \right\}.
  \end{displaymath}
\end{theorem}

\begin{proof}
  The result immediately follows by verifying the condition $\eig(A) \in \mathcal{G}(h)$ in Lemma~\ref{lemma:A=Ahat}.
\end{proof}

Theorem~\ref{thm:identifi-largest-sampl-period} in continuous-time system
identification can be understood by analogy with the \emph{Nyquist-Shannon
  sampling theorem} in signal processing. The \emph{Nyquist-Shannon sampling
  theorem} gives conditions on sampling frequencies, by looking at spectral
information of signals, under which continuous signals can be uniquely
reconstructed from their discrete-time signals. As an analogy,
Theorem~\ref{thm:identifi-largest-sampl-period} addresses that continuous-time
LTI systems can be uniquely reconstructed from their discrete-time systems under
a condition that is built based on the spectral information of the $A$ matrix.

Now we would like to show a property of matrix exponential and logarithm, which
further leads to a test criterion on system aliasing.  See
Appendix~\ref{appdix:test-criteria-sys-alias} for the proofs of
Lemma~\ref{lemma:diff-A-h1h2} and
Proposition~\ref{prop:pred-error-diff-distribution}.
\begin{lemma}
  \label{lemma:diff-A-h1h2}
  Considering $h_1, h_2 \in \mathbb{R}_+$ and $A \in \mathbb{R}^{n \times n}$, let
  $\hat{A}$ be defined by $\hat{A} = \Log(\exp(h_1 A))/h_1$. Then $\exp(h_2 A) =
  \exp(h_2 \hat{A})$ if and only if $h_2/h_1 \in \mathbb{N}$ or $A = \hat{A}$.
\end{lemma}

\begin{proposition}
  \label{prop:pred-error-diff-distribution}
  Consider the dynamical system \eqref{eq:dyna-sys-ss-cont} without inputs (i.e.
  $B = 0$), and two sampling periods $h_1, h_2 \in \mathbb{R}_+$ such that
  $h_2/h_1 \notin \mathbb{N}$.  Let $\hat{A} = \Log(\exp(h_1A))/h_1$ and
  $\hat{A} \neq A$.  The one-step prediction errors w.r.t. $h_2$ are defined as
  \begin{align*}
    \epsilon(t_k) &= x(t_k) - \exp(h_2 A) x(t_{k-1}),\\
    \hat{\epsilon}(t_k) &= x(t_k) - \exp(h_2 \hat{A}) x(t_{k-1}).
  \end{align*}
  Assuming that $\mathbb{E}(x(t_{k-1})) \neq 0$, it yields
  \begin{equation*}
    \mathbb{E}(\epsilon(t_k)) = 0, \quad
    \mathbb{E}(\hat{\epsilon}(t_k)) \neq 0.
  \end{equation*}
\end{proposition}

% \begin{remark}
%   In Proposition~\ref{prop:pred-error-diff-distribution}, if
%   $\mathbb{E}(x(t_{k-1})) = 0$, then
%   $\mathbb{E}(\epsilon(t_k)) = \mathbb{E}(\hat{\epsilon}(t_k)) = 0$. However, we
%   still have
%   $\operatorname{var}(\epsilon(t_k)) \neq \operatorname{var}(\hat{\epsilon}(t_k))$
%   (at least shown by examples in numerical computation). Unfortunately it is
%   particularly difficult to prove in theory due to the integration
%   \eqref{eq:disc-time-sys-wn-cov}, which is computed numerically.
% \end{remark}

We have similar results for the case with inputs, as stated in
Proposition~\ref{prop:pred-error-diff-distribution-B} in
Appendix~\ref{appdix:test-criteria-sys-alias}, where we no longer require
$\mathbb{E}(x(t_k)) \neq 0$ due to the benefits from inputs. Meanwhile, according to
the condition \eqref{eq:pred-err-diff-0-B-condition}, it is possible that a carefully
designed input signal invalidates the test criterion that is built by evaluating
$\mathbb{E}(\hat{\epsilon}(t_k))$, which in practice may not be a problem.
The results in Proposition~\ref{prop:pred-error-diff-distribution} and
Proposition~\ref{prop:pred-error-diff-distribution-B} can be understood by
Figure~\ref{fig:samples-diff-h1h2}, where the output prediction of $\hat{A}$ (that is
estimated from samples in $h_1$) presents different values from that of $A$ in
another sampling period $h_2$ and it results in that the expectation of one-step
prediction errors is no longer zero.  The test criterion on system aliasing is
summarized as follows:
\begin{figure}[htb]   %htbp!H
  \centering
  \includegraphics[width=.45\textwidth]{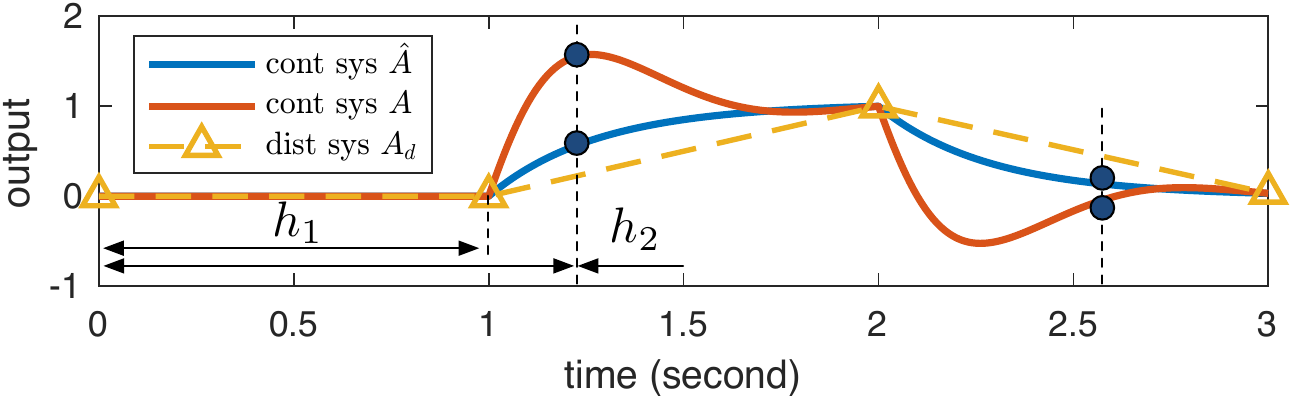}
  \caption{System responses of the dynamical system \eqref{eq:dyna-sys-ss-cont} with
    $A, \hat{A}$, where $A_d = \exp(h_1A), \hat{A} = \Log(A_d)/h_1$, and the input
    signal is the square wave with period $2h_1$. The dots in deep
    blue are samples with the sampling period $h_2$.}
  \label{fig:samples-diff-h1h2}
\end{figure}

\begin{criteria}[system aliasing]
  Identify $A, B$ by PEM or ML (denoting the estimates by $\hat{A}, \hat{B}$)
  assuming no system aliasing under the sampling period $h_1$, i.e. $\hat{A}$
  asymptotically converges to $\Log(\exp(h_1A))$.  Choose another sample period $h_2$
  such that $h_2/h_1 \notin \mathbb{N}_+$, and sample by $h_2$ the system responses
  with non-zero initial conditions or non-zero inputs (assuming
  \eqref{eq:pred-err-diff-0-B-condition} is satisfied). Use $\hat{A}, \hat{B}$ to
  calculate the one-step prediction errors $\{\epsilon(t_k)\}$. Perform \emph{t-test} to
  obtain the \emph{p-value} to make decisions, where the null hypothesis is that
  $\{\epsilon(t_k)\}$ comes from a normal distribution with mean zero and unknown
  variance. Rejecting the null hypothesis implies the existence of system aliasing.
\end{criteria}

\begin{figure}[htbp]   %htbp!H
  \centering
  \includegraphics[width=.5\textwidth]{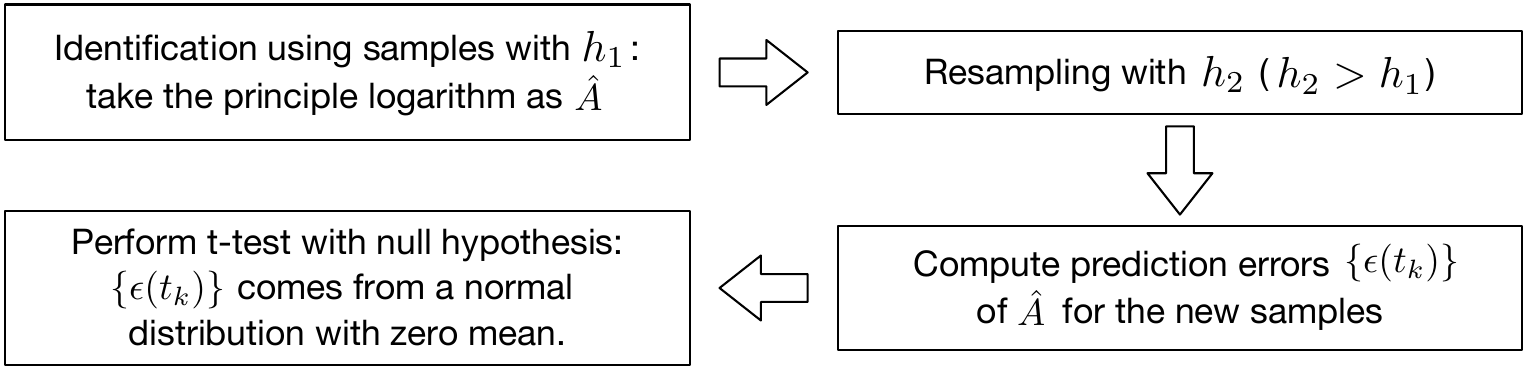}
  \caption{A chart flow of the test criterion on system aliasing.}
  \label{fig:chartflow-test-criteria}
\end{figure}

\section{No system aliasing: sparse network reconstruction}
\label{sec:methods}

Considering the systems given by ~\eqref{eq:dyna-sys-ss-cont} and
\eqref{eq:dyna-sys-ss-output},
the likelihood function is determined by the multiplication rule for conditional
probability\cite{Astrom1980}
\begin{math}
  p(Y^N | \vartheta) \triangleq p(y(t_N), \dots, y(t_1) | \vartheta) =
  p(y(t_N) | y(t_{N-1}),\vartheta)\,
  p(y(t_{N-1}) | y(t_{N-2}),\vartheta)\,
  \cdots p(y(t_1)| \vartheta),
\end{math}
% \begin{equation*}
%   % \label{eq:likelihood-func}
%   \begin{array}{l@{\;}l@{\;}l}
%     L(\vartheta) &= &p(y(t_N) | \vartheta) \\
%               &= &p(y(t_N) | y(t_{N-1}),\vartheta)\, p(y(t_{N-1}) | y(t_{N-2}),\vartheta)\,
%                 \cdots p(y(t_1))
%               & &p(y(t_1) | y(t_0),\vartheta)\, p(y(t_0)),
%   \end{array}
% \end{equation*}
where $\vartheta$ denotes the parameters under estimation, which parameterizes
$A, B, R, m_0, R_0$.
With the assumption that $w(t), x(t_0)$ are jointly Gaussian,
the negative logarithmic likelihood function is
\begin{equation}
  \label{eq:likelihood-func}
  \begin{array}{@{}l@{\;}l@{\;}l}
    L(\vartheta) &= -2\log p(Y^N|\vartheta)
                   = \sum_{k=1}^N \log \det \Lambda(t_k,\vartheta)\\
    &+ \sum_{k=1}^N \epsilon^T(t_k,\vartheta) \Lambda^{-1}(t_k,\vartheta)
                   \epsilon(t_k,\vartheta) + \mathrm{const},
    % \triangleq L_1(\vartheta) + L_2(m_0, R_0)\\
    % &=&\sum_{k=1}^N \epsilon^T(t_k,\vartheta) \Lambda^{-1}(t_k,\vartheta) \epsilon(t_k,\vartheta)
    %   + \sum_{k=1}^N \log \det \Lambda(t_k,\vartheta)\\
    % & &+ (x_0 - m_0)^T R_0^{-1} (x_0 - m_0) + \log\det R_0
    %   + \mathrm{const},
  \end{array}
\end{equation}
where $\epsilon(t_k,\vartheta) \coloneqq y(t_k) - \hat{y}(t_k | t_{k-1},\vartheta)$,
$\hat{y}(t_k | t_{k-1},\vartheta)$ denotes the conditional mean of $y(t_k)$, and
$\Lambda(t_k,\vartheta)$ the corresponding covariance matrix.  The optimal prediction of
$y(t_k)$ (i.e. $\hat{y}(t_k), \Lambda(t_k)$) is obtained using Kalman filters
(e.g. \cite{Astrom1980,Ljung2010}),
\begin{equation}
  \label{eq:pred-Kalman-filter}
  \begin{array}{r@{\;}l}
    \hat{y}(t_k|t_{k-1}) &= C \hat{x}(t_k|t_{k-1}) \\[.5mm]
    \hat{x}(t_k | t_k) &= \hat{x}(t_k|t_{k-1}) + K(t_k) \epsilon(t_k) \\
    \displaystyle\frac{\mathrm{d}}{\mathrm{d}t} \hat{x}(t|t_k) &= A \hat{x}(t|t_k) +
              B u(t), \quad t_k \leq t \leq t_{k+1}\\
    % \hat{x}(t_{k+1}|t_k) &= A_d \hat{x}(t_k|t_k) + B_d u(t_k)\\
    K(t_k) &= P(t_k|t_{k-1}) C^T \Lambda^{-1}(t_k) \\[.5mm]
    P(t_k|t_k) &= P(t_k|t_{k-1}) - K(t_k) C P(t_k|t_{k-1})\\
    \displaystyle \frac{\mathrm{d}}{\mathrm{d}t}P(t|t_k) &= A P(t|t_k) + P(t|t_k)A^T
               + R, \quad t_k \leq t \leq t_{k+1}\\
    % P(t_{k+1}|t_k) &= A_d P(t_k|t_k) A_d^T + R_{d}\\
    \Lambda(t_k) &= CP(t_k|t_{k-1})C^T,
  \end{array}
\end{equation}
where the initial condition is $\hat{x}(t_1|t_{0}) = m_0,\ P(t_1|t_{0}) = R_0$.
% Under certain fair conditions (e.g. \cite{Bertsekas1995}), $\{P(t_k)\}$ in
% \eqref{eq:pred-Kalman-filter} converges, which allows us to use the steady-state
% Kalman filter
% \begin{equation}
%   \label{eq:steady-state-Kalman-filter}
%   \begin{array}{l@{\;}l}
%     P &= A_d (P - P(P+R_2)^{-1}P)A_d^T + R_1,\\
%     K &= P C^T R^{-1},\\
%     R &= R_2 + CPC^T.
%   \end{array}
% \end{equation}
% Therefore, in our setup, the variables $K, R$ are considered to be constant, which is
% the case widely used, e.g. \cite[Sec.~3.6]{Astrom1980},\cite{Ljung2010,Ljung1998}.
Considering the equidistant sampling and assuming the input is constant over the
sampling periods, the matrix $\Lambda$ and $K$ appears in
\eqref{eq:pred-Kalman-filter} can be treated as constant matrices by using
steady-state Kalman filtering\cite[Sec.~3.6]{Astrom1980}.
% In system identification, the key step is to iteratively compute the prediction errors and the corresponding gradient.

Now consider the full-state measurement case (i.e. $C = I$) and restrict the noise to
process noise. The calculation of prediction $\hat{y}(t_k|t_{k-1})$ becomes
particularly simple since $K$ in \eqref{eq:pred-Kalman-filter} always equals the
identity, which yields
\begin{equation}
  \label{eq:special-case-pred-y}
  \begin{array}{r@{\;}l}
    \epsilon(t_k, \vartheta) &= y(t_k) - A_d y(t_{k-1}) - B_d u(t_{k-1}),\\
    \Lambda(t_k, \vartheta) &= R_d,
  %   \hat{y}(t_k|t_{k-1}) &= \hat{x}(t_k|t_{k-1}),\\
  %   \hat{x}(t_k|t_k) &= \hat{x}(t_k|t_{k-1}) + \epsilon(t_k) = y(t_k),\\
  %   \hat{x}(t_{k+1}|t_k) &= A_d \hat{x}(t_k|t_k) + B_d u(t_k),
  \end{array}
\end{equation}
where $A_d, B_d, R_d$ are defined via $A,B,R$ in \eqref{eq:formula-Ad-Bd}, \eqref{eq:disc-time-sys-wn-cov}.
% It is easy to see that the minimization of $L(\vartheta)$ can be done separately in
% terms of $A, B, R$ and $m_0, R_0$. In the current case we cannot estimate the
% distribution of $x_0$ better than a delta function, which makes little sense to
% include two terms of $m_0, R_0$ in \eqref{eq:likelihood-func}. Alternative the
% likelihood without $L_2(m_0, R_0)$ can be obtained by treating $x_0$ to be the
% deterministic, where $p(y(t_1) | y(t_0), \vartheta)p(y(t_0)|\theta)$ simplifies into $p(y(t_1)| \vartheta)$.
Here we resolve $p(y(t_1)|\vartheta)$ by using $p(y(t_1)|y(t_0), \vartheta)$,
where $y(t_0)$ is treated to be the deterministic and hence is removed from the
conditional variables, and takes the first sample as its value. This
simplification is due to the fact that $K \equiv I, P(t_k|t_k) \equiv 0$ and the
measurement of $x_0$ is available (using $y(t_0)$), which also leads to that the
best estimation of the distribution of $x_0$ is nothing better than a delta
function (even if including the probability assumption of $x_0$ in maximum
likelihood, i.e.
$p(y(t_1)|\vartheta) = p(y(t_1)|y(t_0), \vartheta) p(y(t_0)|\vartheta)$ and
$p(y(t_0)|\vartheta)$ takes the Gaussian density with mean $m_0$ and covariance
matrix $R_0$). Alternatively, the likelihood function \eqref{eq:likelihood-func}
can be obtained directly by considering \eqref{eq:dyna-sys-ss-discr} without
using Kalman filtering. However, the above standard procedure values when we
deal with general cases (i.e. $C \neq I$).  Noticing the particular
parameterization \cite[p.~92,~206]{Ljung1998}, maximizing likelihood can be
performed as follows\footnote{Similar to \cite[p.~219]{Ljung1998}, one instead
  firstly minimizes the cost function analytically with respect to $\vartheta$
  for every fixed $R$. Due to the particular parameterization, the resultant
  optimization no longer depends on $R$.}:
\begin{subequations}
  \label{eq:ML-2steps}
  \begin{align}
    \label{eq:ML-2steps-PEM}
    \hat{\theta} &= \textstyle\operatorname{argmin}_{\theta} \sum_{k=1}^N \epsilon(t_k,\theta)^T \epsilon(t_k,\theta),\\
    \label{eq:ML-2steps-R}
    \hat{R}_d &= \textstyle\frac{1}{N} \sum_{k=1}^N
\epsilon(t_k,\hat{\theta}) \epsilon^T(t_k, \hat{\theta}),
  \end{align}
\end{subequations}
where $\theta$ is composed of $A,B$.
% Note that this is not generally true for maximizing likelihood, which should be
% solved by \cite[p.~219]{Ljung1998} in general for multivariable cases.
To estimate $\theta$, instead of minimizing the prediction error as
\eqref{eq:ML-2steps-PEM}, we impose the $l_1$-penalty to favor the sparse
solution in network reconstruction. This is due to the observations in
Section~\ref{subsec:ambig-matrix-log}: the consistency of ML may fail to present
us with a correct network structure unless appropriate thresholds of zero for
each row of $A, B$ are selected, which is hardly implemented in practice.

\begin{remark}
  \label{rmk:challenge-from-measurement-noise}
  If we include measurement noise\footnote{Assume that it is reasonable to
    determine the network by $A, B$ similarly, even though we don't have network
    models well-defined from the state-space representations with measurement
    noise.}  or consider the output measurement case $C\neq I$, the prediction
  includes the Kalman filter gain $K$, which depends on $A, R$ and the
  covariance of measurement noise.  It deserves to be emphasized that, due to
  the possible large sampling periods $h$, the numerical tricks used in
  \cite{Astrom1980,Ljung2010} may no longer be valid to compute the gradient of
  the prediction error. We have to analytically calculate the gradient as far as
  possible until the numerical computation is no longer restricted by $h$. This
  problem becomes fairly complicated.
\end{remark}

\subsection{The cost function in matrix forms and the gradients}
\label{sec:pred-errors-mat-gradient}

The reconstruction algorithm is supposed to infer a sparse network, i.e. $A$ is
sparse.  Due to the nonlinear least-square cost function \eqref{eq:ML-2steps-PEM}, it
no longer satisfies the setup of Sparse Bayesian Learning proposed in
\cite{Tipping2001}. Here we enhance sparsity by heuristically imposing the $l_1$-norm
of $A$ as the penalty to the PEM cost function as the first tentative treatment.

Considering the measurement signal $Y^N$, let
\begin{equation*}
  \begin{array}{l@{\;}l}
    X_+ &\triangleq [ y(t_1),\; y(t_2),\; \dots,\; y(t_N) ],\\
    X_- &\triangleq [ y(t_0),\; y(t_1),\; \dots,\; y(t_{N-1}) ],\\
    U_- &\triangleq [ u(t_0),\; u(t_1),\; \dots,\; u(t_{N-1}) ],
  \end{array}
\end{equation*}
where $X_+,X_- \in \mathbb{R}^{n \times N}$.
The matrix form of the $l_1$-regularised PEM problem is formulated as
\begin{equation}
  \label{eq:optim-probl-type-2}
  \begin{array}{l@{\;}l}
    \minimize{A} &  \big\| X_+ - \exp(hA) X_- - \displaystyle\int_0^h
                   \exp(sA)\,\mathrm{d}s\, B U_- \big\|_F^2 \\
                 &+ \lambda \left\| A \right\|_1,
  \end{array}
\end{equation}
where $\lambda \in \mathbb{R}^+$ and $h \in \mathbb{R}^+$ is the fixed and known
sampling period, and the $l_1$ norm $\| A \|_1 \coloneqq \sum_{i,j=1}^n |A_{ij}|$
($A_{ij}$ denotes the $(i,j)$-th element of $A$).  To avoid dealing with tensors, we
use the vectorized form of (\ref{eq:optim-probl-type-2}) as follows:
\begin{equation}
  \label{eq:optim-probl-kron-type-2}
  \begin{array}{l@{\;}l}
    \minimize{A} &  \big\| \kvec(X_+) - (X_-^{T} \otimes I_n) \kvec(\exp(hA)) \\
                 &-(U_-^TB^T \otimes I_n) \displaystyle\int_0^h \kvec(\exp(sA))\,\mathrm{d}s \big\|_2^2\\
                 &+ \lambda \left\| \kvec(A) \right\|_1,
  \end{array}
\end{equation}
where $\kvec (X_+) \in \mathbb{R}^{nN}$, $\kvec (\exp(A h)) \in \mathbb{R}^{n^2}$,
$(X_-^{T} \otimes I_n) \in \mathbb{R}^{nN \times n^2}$, and $\otimes$ is the
\emph{Kronecker product}.

The problem \eqref{eq:optim-probl-type-2} is challenging in optimization by noticing
that it is: non-convex due to matrix exponential; not globally Lipschitz; and
non-differentiable. The intuitive idea here is to use the the Gauss-Newton framework,
in which each iteration is to solve a constrained $l_1$-regularized linear least
square problem. Let
\begin{equation}
  \label{eq:obj-func-r}
  \begin{array}{l@{\;}l}
    r(A,B) \triangleq &\kvec(X_+) - (X_-^{T} \otimes I_n) \kvec(\exp(hA)) \\
                    &-(U_-^TB^T \otimes I_n) \displaystyle\int_0^h \kvec(\exp(sA))\,\mathrm{d}s,
  \end{array}
\end{equation}
\(\phi(A,B) \coloneqq r(A,B)^T r(A,B)\), and $f(A,B) \triangleq \phi(A,B) + \lambda
\|\kvec(A))\|_1$, which is the objective function of
\eqref{eq:optim-probl-kron-type-2}. Then \(\min_A \phi(A,B)\) denotes the problem
\eqref{eq:optim-probl-kron-type-2} without $l_1$-penalisation. The gradient of
\(\phi(A,B)\) w.r.t. $A$ is
\begin{math}
  % \label{eq:gradient-phi-A}
  \nabla_A \phi = 2 J_A^T(A,B) r(A,B),
\end{math}
where
\begin{equation}
  \label{eq:jacobian-J-A}
  J_A = -h (X_-^T \otimes I_n) K(hA) -(U_-^TB^T \otimes I_n) \displaystyle\int_0^h sK(sA)\,\mathrm{d}s,
\end{equation}
and \(K(A)\) is defined in Theorem~\ref{thm:kron-repr-frechet-deriv}. The function
$sK(sA), s\!\in\![0,h]$ is integrable by noticing $\|sK(sA)\| \leq h \exp(h\|A\|)$
($\|\cdot\|$ denotes any matrix norm). The matrix function $K(hA)$ and the
integration can be calculated numerically given $A$.  To compute the gradient of
$\phi(A,B)$ w.r.t. $B$, using the matrix identity $\kvec(AXB) = (B^T \otimes A) \kvec(X)$
again for the term of $U$ in \eqref{eq:optim-probl-type-2}, it yields
\begin{math}
  % \label{eq:kron-vec-id-U}
    \big(U_-^TB^T \otimes I_n\big) \int_0^h \kvec(\exp(sA))\,\mathrm{d}s
     =
    \big(U_-^T \otimes \int_0^h \kvec(\exp(sA))\,\mathrm{d}s \big) \kvec(B).
\end{math}
Then it follows that
\begin{math}
  % \label{eq:gradient-phi-B}
  \nabla_B \phi = 2 J_B^T(A,B) r(A,B),
\end{math}
where
\begin{equation}
  \label{eq:jacobian-J-B}
  J_B = U_-^T \otimes \displaystyle\int_0^h \kvec(\exp(sA))\,\mathrm{d}s.
\end{equation}
If we assume $B$ is diagonal and $\nabla_B\phi$ is calculated w.r.t. each diagonal
element of $B$, then $\nabla_B\phi = 2 \big((I \otimes \mathbf{1})J_B\big)^T r(A,B)$,
where $I$ is an $n \!\times\! n$ identity matrix and $\mathbf{1}$ is an
$n^2$-dimensional row vector of 1's.  In a sum, the gradient of $\phi(A,B)$ is
\begin{equation}
  \label{eq:gradient-phi}
  \nabla \phi =
  \begin{bmatrix}
    \nabla_A \phi \\ \nabla_B \phi
  \end{bmatrix}
  = 2
  \begin{bmatrix}
    J_A^T \\ J_B^T
  \end{bmatrix} r(A,B)
  \triangleq 2 J^T(A,B) r(A,B).
\end{equation}
% where
% \begin{equation}
%   \label{eq:jacobian-J}
%   J(A,B) =
%   \begin{bmatrix}
%     J_A & J_B
%   \end{bmatrix}.
% \end{equation}

\subsection{A special case: update A with fixed $B$}
\label{subsec:special-case:-update-A}

The \emph{subspace method} in system identification presents us with nice
initial estimation of $A_0, B_0$ (e.g. see \cite{Viberg2002}). Concerning the
task of network reconstruction, we would like to infer a sparse $A$ from
data. As a special case, we only update $A$ by solving
\eqref{eq:optim-probl-type-2} with $B$ fixed to be $B_0$. For simplicity, in
this subsection, let
$r(A, B_0) \triangleq r(A), J_A(A, B_0) \triangleq J(A), \phi(A, B_0) \triangleq
\phi(A)$ and $f(A,B_0) \triangleq f(A)$.

A linear approximation of $r(A)$ in a neighbourhood of a given point $A_c$ is
\begin{math}
  % \label{eq:linear-approx-obj-func-r}
  r(A_c) + J(A_c)\kvec(A - A_c).
\end{math}
One may then use this approximation and formulate a $l_1$-regularized linear least squares problem
\begin{equation}
  \label{eq:lsq-approx-opti-probl-incomplete}
  \minimize{A}\; \|r(A_c) + J(A_c) \kvec(A - A_c)\|_2^2 + \lambda \|\kvec(A)\|_1,
\end{equation}
which can be solved to obtain an approximate solution to \eqref{eq:optim-probl-kron-type-2}.
Resolving it in an iterative way amounts to a Gauss-Newton method. However, \(\kvec(A - A_c)\) is not necessary to be a \emph{descent direction} of \eqref{eq:optim-probl-kron-type-2}.

In the $k$-th iteration, to guarantee the step $p_k = \kvec(A-A_k)$ being a descent direction of \eqref{eq:optim-probl-kron-type-2}, the search direction $p_k$ is instead computed from the following constrained optimization problem
\begin{equation*}
  % \label{eq:lsq-approx-opti-probl}
  (P_1) \left\{
  \begin{array}{@{\,}l@{\;\;}l}
    \minimize{p_k \in \mathbb{R}^{n^2}} &\|r(A_k) + J(A_k) p_k\|_2^2 + \lambda \|\kvec(A_k) + p_k\|_1,\\
    \st & \sup_{g \in \partial f(A_k)} g^T p_k \leq 0,
  \end{array}
  \right.
\end{equation*}
where $\partial f(A_k)$ denotes the subdifferential of $f(A)$ at $A_k$, defined as
\begin{equation}
  \label{eq:subgradient-f-Ak}
  \partial f(A_k) \coloneqq \{ \nabla \phi(A_k) + \lambda z : z \in J_1 \times \cdots J_{n^2} \},
\end{equation}
\begin{equation}
  \label{eq:subgradient-Ji}
  J_i \coloneqq
  \begin{cases}
    [-1, 1] &  \text{if } \kvec(A_k)_i = 0 \\
    \{1\}   &  \text{if } \kvec(A_k)_i > 0 \\
    \{-1\}  &  \text{if } \kvec(A_k)_i < 0
  \end{cases},
\end{equation}
in which $\kvec(A_k)_i$ denotes the $i$-th element of $\kvec(A_k)$.
One may have noticed that the constraint in the problem $(P_1)$ is the definition of \emph{descent direction} for $f$ at $A_k$, except replacing $< 0$ with $\leq -\epsilon$ to guarantee the existence of minimum. The problem $(P_1)$ is a convex optimization problem by noticing that $\sup_{g \in \partial f(A_k)} g^T p_k$ is a convex function, which is a \emph{pointwise supremum} over an infinite set of a linear function \cite[chap.~3]{Boyd2004}.
To solve the problem $(P_1)$, we need to explore the constraint and derive an equivalent form (see Appendix~\ref{appdix:derivation-Pprim-P} for details), given as follows.
\begin{equation*}
  (P_1') \left\{
  \begin{array}{@{\,}l@{\;\;}l}
    \minimize{p_k \in \mathbb{R}^{n^2}} &\|r(A_k) + J(A_k) p_k\|_2^2 + \lambda \|\kvec(A_k) + p_k\|_1,\\
    \st & \bar{g}(A_k)^T p_k + \lambda \|W(A_k) p_k\|_1 \leq 0,
  \end{array}
  \right.
\end{equation*}
where
\begin{equation}
  \label{eq:formula-gbar-W}
  \begin{array}{l@{\;}l}
    \bar{g}(A_k) &= \nabla \phi(A_k) + \lambda \sgn(A_k),\\
    W(A_k) &= I - \diag(|\sgn(A_k)|),
  \end{array}
\end{equation}
the identify matrix $I$ is of a compatible dimension, $|\cdot|$ denotes the element-wise absolute value, $\diag(v)$ denotes the diagonal matrix built from vector $v$, and the $\sgn$ function for vectors and matrices is extended from the standard signum function for real numbers, defined as follows: when $x \in \mathbb{R}^n$, $\sgn(x)$ denotes a $n$-dimensional vector whose $i$-th element equals $\sgn(x_i)$; and when $X \in \mathbb{R}^{m \times n}$, $\sgn(X) \coloneqq \sgn(\kvec(X))$.
Now the problem $(P')$ can be easily modeled using CVX in MATLAB and solved by
standard optimization solvers \cite{cvx-manual}.

The iterate is updated via
\begin{equation}
  \label{eq:update-rule-plus-line-search}
  \kvec(A_{k+1}) = \kvec(A_k) + s_k p_k,
\end{equation}
where the step length $s_k$ is determined by \emph{backtracking line search}.
Let $f'(A_k; p_k)$ denote the \emph{directional derivative} of $f$ at $A_k$ in the direction $p_k$, by subcalculus,
\begin{equation}
  \label{eq:directional-derivative}
  \begin{array}{l@{\;}l}
    f'(A_k; p_k) &\coloneqq \sup_{g \in \partial f(A_k)} g^T p_k\\[4pt]
                 &= \bar{g}(A_k)^T p_k + \lambda \|W(A_k) p_k\|_1.
  \end{array}
\end{equation}
Given $\alpha \in (0, 0.5), \beta \in (0,1)$ and an initial value $s_k = 1$, the line search is to perform $s_k \leftarrow \beta s_k$ until
\begin{equation}
  \label{eq:line-search}
  f \big(A_k + \ikvec(s_kp_k)\big) \leq f(A_k) + \alpha s_k f'(A_k; p_k).
\end{equation}
The whole iterative method for \eqref{eq:optim-probl-kron-type-2} is summarized
in Algorithm~\ref{alg:iter-algor-optimisation}. One has to note that this
algorithm may not guarantee that the iterate will converges to the stationary
point. It is lucky that we have good initial values of $A_0, B_0$ to start with
that is provided by the \emph{subspace method} in system identification. Solving
\eqref{eq:optim-probl-type-2} is to search a sparse $A$ in the neighborhood of
$A_0$.  Moreover, we have the following propositions to guarantee fair
properties of this algorithm. See Appendix~\ref{appdix:derivation-Pprim-P} for
the proofs.
\begin{proposition}
  \label{prop:no-zero-p-before-local-optima}
  Let $\hat{f}(A_k, p_k)$ denote the objective function of $(P_1)$ and $p_k^*$ be its
  optimal point. If $p_k^* \neq 0$ and $\sup_{g\in \partial f(A_k)} g^T p_k^* = 0$, then
  $0 \in \partial f(A_k)$.
\end{proposition}

\begin{proposition}
  \label{prop:zero-p-guarantee-local-optima}
  Let $\hat{f}(A_k, p_k)$ and $p_k^*$ be defined in
  Proposition~\ref{prop:no-zero-p-before-local-optima}.
  If $p_k^* = 0$ and $0 \in \operatorname{argmin}_{p_k} \hat{f}(A_k,p_k)$, then
  $0 \in \partial f(A_k)$.
\end{proposition}

Proposition~\ref{prop:no-zero-p-before-local-optima} guarantees that the step $p_k^*$
from solving $(P_1)$ will always be a descent direction of
\eqref{eq:optim-probl-kron-type-2} (i.e. $\sup_{g\in \partial f(A_k)} g^T p_k^* < 0$)
until either it reaches the stationary point or $\{p_k^*\}$ converges to zero. When
$\{p_k^*\}$ approaches to zero, there are two cases: one is
Proposition~\ref{prop:zero-p-guarantee-local-optima} which guarantees that it reaches
the stationary point; the other is described as follows:
\begin{quote}
  $p_k^* = 0$ is the unique optimal point of $(P_1)$ and $0 \notin \operatorname{argmin}_{p_k} \hat{f}(A_k,p_k)$.
\end{quote}
Regarding the second case, indeed, if there exists other optimal point
$p_k^{**} \neq 0$ of $(P_1)$, we instead consider $p_k^{**}$ using
Proposition~\ref{prop:no-zero-p-before-local-optima}. This is why we restrict
$p_k^* =0$ to be the unique optimum of $(P_1)$. In the second case, $\{p_k^*\}$
converges to zero and the objective value $f(A_k)$ also converges. However, in
theory, we fail to prove that the limit point of $\{A_k+ \ikvec(p_k^*)\}$ is a
stationary point. In the sense of applications, it has been fairly good since we are
looking up a sparse solution in the neighborhood of a fairly good estimate.

\begin{algorithm}
  \caption{\small Modified Gauss-Newton for $l_1$-regularized nonlinear least square problems}
  \label{alg:iter-algor-optimisation}
  \small
  \begin{algorithmic}[1]
    \State \textbf{given} $\epsilon >0$, tolerance $\delta > 0$; $\alpha \in (0,0.5)$, $\beta \in (0,1)$
    \State \textbf{initialize} $A_0,B_0$ by the \emph{subspace method}.
    \Repeat
    \State Calculate \(r(A_{k}), J(A_{k}), \nabla \phi(A_{k})\) using \eqref{eq:obj-func-r}, \eqref{eq:jacobian-J-A}.
    \State Calculate \(\bar{g}(A_k), W(A_k)\) using \eqref{eq:formula-gbar-W}.
    \State Compute the search direction \(p_k\) by solving problem \((P')\).
    \State Compute the directional derivative \(f'(A_k; p_k)\) using~\eqref{eq:directional-derivative}.
    \State Determine the step size \(s_k\) by backtracking line search \eqref{eq:line-search}.
    \State Update the iterate \(A_{k}\) by \eqref{eq:update-rule-plus-line-search}.
    \Until{$\|A_k - A_{k+1}\|_2 < \delta$}  % $f(A_k) - f(A_{k+1}) < \delta$
  \end{algorithmic}
\end{algorithm}

\begin{remark}
  The proposed method can be considered as a variant of the \emph{damped Gauss-Newton} method.
  % \footnote{\label{fnt:damped-gauss-newton}It exactly follows the damped Gauss-Newton method, if removing the penalty term.}.
  If the Jacobian matrix \(J(A_k)\) does not have full column rank, one could
  adopt the \emph{Levenberg-Marquardt} method
% \footnote{See Footnote~\ref{fnt:damped-gauss-newton}.}
  and solve
  \begin{equation*}
    % \label{eq:lsq-approx-opti-probl-leveg-marqt}
    \hspace*{-1.3mm}
    \scalebox{0.9}{\(
      \begin{array}{l@{\:\,}l}
        \minimize{p_k \in \mathbb{R}^{n^2}} &\|r(A_k) + J(A_k) p_k\|_2^2 + \mu_k \|p_k\|_2^2 + \lambda \|\kvec(A_k) + p_k\|_1,\\
        \st & \sup_{g \in \partial f(A_k)} g^T p_k \leq -\epsilon.
      \end{array}
      \)}
  \end{equation*}
  % Moreover, if the matrices \(\{J(A_k)^T J(A_k)\}\) are not uniformly bounded or well conditioned, the damped Gauss-Newton may not work effectively.
\end{remark}

\subsection{General cases: update both $A$ and $B$}
\label{subsec:general-case:-update-AB}

Considering the vectorized form \eqref{eq:optim-probl-kron-type-2}, let $\theta$
denote the optimal variables, i.e.
$\theta \triangleq [\kvec(A)^T\; \kvec(\diag(B))^T]^T$, where $\diag(B)$ denotes the
diagonal elements of $B$. The other notations follow that
$r(A,B) \triangleq r(\theta), J(A,B) \triangleq J(\theta), \phi(A,B) \triangleq
\phi(\theta)$ and $f(A,B) \triangleq f(\theta)$. In the same way as
Section~\ref{subsec:special-case:-update-A}, the approximated $l_1$-regularized
linear least square problem with constraints is written as
\begin{equation*}
  (P_2) \left\{
  \begin{array}{@{\,}l@{\;\;}l}
    \minimize{p_k \in \mathbb{R}^{2n^2}} &\|r(\theta_k) + J(\theta_k) p_k\|_2^2 + \lambda
                                           \|\kvec(A_k) + \Lambda p_k\|_1,\\
    \st & \sup_{g \in \partial f(\theta_k)} g^T p_k \leq 0,
  \end{array}
  \right.
\end{equation*}
where $\Lambda = [I\;\: 0]$ is of dimension $n^2\!\times\! 2n^2$, $\partial f(\theta_k)$ denotes the subdifferential of $f(\theta)$ at $\theta_k$, defined as
\begin{equation*}
  \partial f(\theta_k) \coloneqq \{ \nabla \phi(\theta_k) + \lambda \Lambda^T z : z \in J_1 \times \cdots J_{n^2} \},
\end{equation*}
% \begin{equation*}
%   J_i \coloneqq
%   \begin{cases}
%     [-1, 1] &  \text{if } \kvec(A_k)_i = 0 \\
%     \{1\}   &  \text{if } \kvec(A_k)_i > 0 \\
%     \{-1\}  &  \text{if } \kvec(A_k)_i < 0
%   \end{cases},
% \end{equation*}
and $J_i, i = 1,\dots,n^2$ are defined in \eqref{eq:subgradient-Ji}.
Equivalently, we solve the following convex-constrained convex problem to update $A, B$
by $\theta_{k+1} = \theta_{k} + p_k$.
\begin{equation*}
  (P_2') \left\{
  \begin{array}{@{\,}l@{\;\;}l}
    \minimize{p_k \in \mathbb{R}^{2n^2}} &\|r(\theta_k) + J(\theta_k) p_k\|_2^2 + \lambda
                                           \|\kvec(A_k) + \Lambda p_k\|_1,\\
    \st & \bar{g}^T p_k + \lambda \|W(A_k) \Lambda p_k\|_1 \leq 0,
  \end{array}
  \right.
\end{equation*}
where $\bar{g} \triangleq \nabla\phi(\theta) + \lambda \Lambda^T \sgn(A_k)$ and $W(A_k) = I - \diag(|\sgn(A_k)|)$.
The backtracking line search is equipped in the same way as
Section~\ref{subsec:special-case:-update-A} and the algorithm trivially follows by
modifying Algorithm~\ref{alg:iter-algor-optimisation}.

% \section{System aliasing: search over branches}
\section{System aliasing and bounded constraints}
\label{sec:syst-alias-bound-constraints}

In the previous section we hinted that the conditions for no \emph{system aliasing}
follow as a consequence of bounded eigenvalues.  In this section we follow this path
and study the problem in the presence of \emph{system aliases}.

Consider the case of system aliasing, i.e. $h$ is NOT chosen small enough such that
$\mathscr{E}(A, I, h, \mathscr{S}_A) = \{A\}$. In order to find out $A$ among the
aliases we need extra information, for instance, the properties of $A$ known \emph{a
  priori}. Here we assume that the ground truth $A$ is the sparest solution in
$\mathscr{E}(A, I, h, \mathscr{S}_\kappa)$ and $\kappa \in \mathbb{R}$ as an upper
bound that has been prescribed. The set $\mathscr{S}_\kappa$ will be defined after
giving Definition~\ref{def:coord-change-norm}.  $A$ can be searched by the criterion
\begin{equation}
  \label{eq:sparsity-multiple-logm-bounded}
  \minimize{\hat{A} \in \mathscr{E}(A, I, h, \mathscr{S}_\kappa)}  \|\hat{A}\|_0.
  \vspace*{-1.5mm}
\end{equation}
Here is a niche that is the calculation of $\mathscr{E}(A,I,h,\mathscr{S}_\kappa)$
from data.
By definition,
\begin{equation}
  \label{eq:def-set-S}
  \mathscr{E}(A, I, h, \mathscr{S}_\kappa) = \big\{\tilde{A} \in \mathscr{S}_\kappa: \exp(h \tilde{A}) = A_d \big\},
\end{equation}
where $A_d = \exp(hA)$.
Even if we know $A_d$ has consistent estimation via PEM or ML, considering
the observations in Section~\ref{subsec:ambig-matrix-log}, we know the workflow,
that is estimating $A_d$ and then obtaining $\mathscr{E}$ by matrix logarithms, is not
robust in the presence of noise.  In this section, we focus on studying the
possibility of searching $A$ in the set of system aliases
$\mathscr{E}(A,I,h,\mathscr{S}(\kappa))$ using the prior information.

\begin{definition}[$Z$-weighted norm]
  \label{def:coord-change-norm}
  Let $h_Z(A) = Z^{-1}AZ$,
  where $Z$ is the matrix defined in Theorem~\ref{thm:matrix-logarithm-classification}.
  Then the norm is defined as $ \|h_Z(\cdot)\|_F = \|\cdot\|_F \circ h_Z$.
\end{definition}
To formulate $\mathscr{S}_\kappa$, we introduce this special norm of ${A}$, which is equivalent to the Frobenius norm up to a change of coordinates.
The matrix $Z$ is constant, which can be obtained by Jordan decomposition of
$A_d$. One can observe that
\begin{equation*}
  \|h_Z(\hat{A})\|_F = \text{vec}(\hat{A})^T(Z^T \otimes Z^{-1})^T(Z^T \otimes Z^{-1})\text{vec}(\hat{A})
\end{equation*}
is a proper $(Z^T \otimes Z^{-1})^T(Z^T \otimes Z^{-1})$-weighted vector norm in
terms of $\kvec(\hat{A})$.  Using $\| h_Z(\cdot) \|_F$ is on the one hand simplifying
the analysis we conduct throughout this section, and on the other explicitly
penalizes the imaginary part of the eigenvalues without ``distorting'' them through
the transformation by $Z$.

\begin{figure}[htb]   %htbp!H
  \centering
  \includegraphics[width=.45\textwidth]{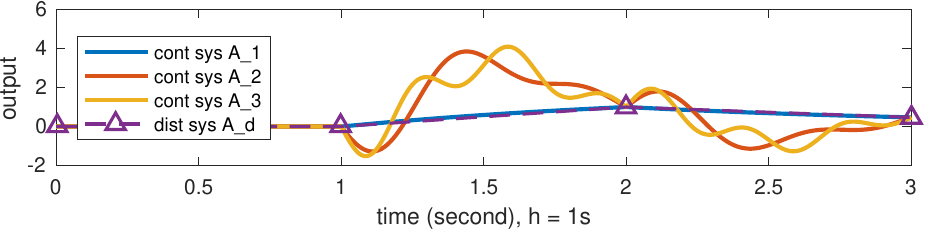}
  \caption{An example of system responses of multiple system aliases, i.e.
    $A_i\: (i \!=\!  1, 2, 3)$ that satisfies $\exp(hA_i) = A_d$, where $A_2$ is the
    ground truth but $A_1$ is the principle logarithm, and the input signal is a square
    wave of period $2h$.}
  \label{fig:responses-system-aliases}
\end{figure}

Now we define $\mathscr{S}_\kappa$ using the norm $\|h_Z(\cdot)\|_F$. The basic idea
is that one should exclude such $A$'s whose imaginary parts of eigenvalues are too
large, which implies their system response will show wild fluctuation, as illustrated
in Figure.~\ref{fig:responses-system-aliases}.
That's why we need to consider a reasonable set $\mathscr{S}_\kappa$ rather than
$\mathbb{R}^{n \times n}$ in \eqref{eq:def-set-S}.  In practice, even if we know
the sampling frequency is not high enough to guarantee no system aliasing, we
could still believe that the measurements do not miss too many fluctuation
between samples.  To make the constraint in
\eqref{eq:sparsity-multiple-logm-bounded} practically meaningful, we restrict
$\mathscr{S}$ to be a norm bounded subset
\begin{equation}
  \label{eq:norm-bounded-S}
  \mathscr{S}_\kappa = \big\{ \tilde{A} \in \mathbb{R}^{n \times n}: \|h_Z(\tilde{A})\|_F \leq \kappa \big\}.
\end{equation}
In the following we will show that the feasible set of
\eqref{eq:sparsity-multiple-logm-bounded} has only finite elements, which
implies it can be solved at least by brute force methods.

Let $M \coloneqq \diag(m_1, m_2, \dots, m_p)$, $\textit{j} \coloneqq [j_1, j_2,
\dots, j_p]$\footnote{We use a different font type from $j$ to avoid misunderstanding
$\textit{j}$ as a scalar variable for indexes.}and $\beta \coloneqq [\beta_1, \beta_2, \dots, \beta_p]$, where   $\log(\lambda_k) \triangleq \alpha_k + i\pi \beta_k$, $k = 1,\dots,p$, and $j_k, \lambda_k$ are defined in Theorem~\ref{thm:matr-logar-Gantmacher}. A function $\mathscr{I}$ is defined as
\begin{equation}
  \mathscr{I}(\textit{j}, \delta) \coloneqq \delta^T M \delta + (2\textit{j} + \beta)^T M \delta,
  \label{eq:func-def-dist-Fro-Aij}
\end{equation}
where $\textit{j}, \delta \in \mathbb{Z}^{p}$. Moreover, it satisfies
$\mathscr{I}(\textit{j}, \delta) = \mathscr{I}(0, \textit{j}+\delta) - \mathscr{I}(0, \textit{j})$, which follows by noticing
\begin{alignat}{1}
  \mathscr{I}(\textit{j}, \delta) = \left(\delta + \textit{j} + {\beta}/{2}\right)^T & M \left(\delta + \textit{j} + {\beta}/{2}\right) \nonumber\\
                       - \left(\textit{j} + {\beta}/{2}\right)^T & M \left(\textit{j} + {\beta}/{2}\right).
                       \label{eq:func-def-dist-Fro-Aij-compl-square-form}
\end{alignat}
Moreover, let $A_0$ denote a special matrix logarithm for which all $j_k\ (k = 1,\dots,p)$ in \eqref{eq:mat-log-all-sols} are equal to $0$.

\begin{definition}[equivalence relations]
  \label{def:equivalence-relation}
  Let $\mathcal{S}$ denote the set of all primary matrix logarithms
  \begin{equation}
    \label{eq:def-set-cal-S}
    \mathcal{S} \coloneqq \big\{\tilde{A} \in \mathbb{R}^{n \times n}: \exp(h \tilde{A}) = A_d \big\}.
  \end{equation}
  An \emph{equivalence} relation ``$\sim$'' is defined on $\mathcal{S}$ as a binary relation:
  for any $A_1,A_2 \in {\mathcal{S}}$, $\textit{j}^{(1)}$ and $\textit{j}^{(2)}$ are defined for $A_1, A_2$, respectively, we say $A_1 \sim A_2$ if
  \begin{math}
    \mathscr{I}(\textit{j}^{(1)}, \textit{j}^{(2)} - \textit{j}^{(1)}) = 0.
  \end{math}
\end{definition}

\begin{lemma}
  \label{lemma:logm-As-condition-equality}
  Let ${\mathcal{S}}$ be the set defined in \eqref{eq:def-set-S} and
parametrized by~\eqref{eq:logm-classifiction-primary} in Theorem~\ref{thm:matrix-logarithm-classification}.
For any $A_1, A_2 \in {\mathcal{S}}$, $\|h_Z(A_1)\|_F = \|h_Z(A_2)\|_F$ if and only if $A_1 \sim A_2$.
\end{lemma}

% \begin{remark}
%  It is not necessary that $A_0$ is the principal matrix logarithm (consider the case when the principal logarithm does not exist), nor does it have to be the logarithm with the smallest (weighted) Frobenius norm.
% \end{remark}

\begin{lemma}
  \label{lemma:finite-elements-equiv-class}
  Given any $\bar{A} \in {\mathcal{S}}$, there exists a finite number of $A_i \in {\mathcal{S}}$ that satisfies $A_i \sim \bar{A}$.
\end{lemma}

\begin{lemma}
  \label{lemma:finite-A-Fro-less-k}
  There exists a finite number of $A_i \in {\mathcal{S}}$ such that $\|h_Z(A_i)\|_F \leq \kappa$.
\end{lemma}

% \begin{remark}
%   We could have a more precise bound of $\delta_i$. Let $\mu(i) \coloneqq \minimize{\delta} \sum_{k \neq i} m_k (\delta_k + \beta_k/2)^2$. Then $\delta_i\ (i = 1,\dots,p)$ should satisfy
%   \begin{equation}
%     \label{eq:delta-bound-k-better-upperbound}
%     |\delta_i + \beta_i/2| \leq \sqrt{\frac{(\beta/2)^TM(\beta/2) + (\kappa^2 - \kappa_0^2) - \mu(i)}{m_i}}.
%   \end{equation}
%   Moreover, we have the solution to $\mu(i), i = 1,\dots,p$:
%   \begin{align}
%     \mu(i) = \min \{\mathscr{I}(0, \delta_{/i}) : &(\delta_{/i})_k = \lceil -\beta_k/2 \rceil \text{ or } \lfloor -\beta_k/2 \rfloor,\nonumber\\
%                                        &k \neq i;\ (\delta_{/i})_i = 0 \}.
%   \label{eq:mu-in-delta-bound}
%   \end{align}
% \end{remark}

\begin{proposition}[lower boundness of logarithms]
  \label{thm:identifi-boundness-A}
  Let ${\mathcal{S}}$ be the set defined in \eqref{eq:def-set-S}.
Given any $\bar{A} \in {\mathcal{S}}$, there exists $M(\bar{A}) > 0$, such that for any $A \in \{A \in {\mathcal{S}} : A \nsim \bar{A} \}$, it holds that
  \begin{equation*}
    \left| \| h_Z(A) \|_F - \| h_Z(\bar{A}) \|_F \right| \geq M.
  \end{equation*}
\end{proposition}

\begin{proposition}
  \label{lemma:strict-opti-uniq-via-boundness}
  Let ${\mathcal{S}}$ be the set defined in \eqref{eq:def-set-S}.
For any $\bar{A} \in {\mathcal{S}}$, there exist $\kappa_l, \kappa_u \in \mathbb{R}$ in $\mathscr{S}(\kappa_l, \kappa_u) = \{\tilde{A} \in \mathbb{R}^{n \times n}: \kappa_l \leq \|h_Z(\tilde{A})\|_F \leq \kappa_u \}$ such that \eqref{eq:sparsity-multiple-logm-bounded} has a unique optimal point in the sense of the equivalence relation in Definition~\ref{def:equivalence-relation}.
\end{proposition}

\begin{proof}
  It immediately follows by choosing
  \begin{align*}
  \kappa_l &> \max\{0, \|h_Z(\bar{A})\|_F - M(\bar{A}) \},\\
  \kappa_u &< \|h_Z(\bar{A})\|_F + M(\bar{A}),
  \end{align*}
  where $M(\bar{A})$ is the lower bound on the gap between $\bar{A}$ and any $A \nsim \bar{A} \in {\mathcal{S}}$, defined in Theorem~\ref{thm:identifi-boundness-A}.
\end{proof}

\section{Numerical examples}
\label{sec:simulations}

This section shows numerical examples of the proposed algorithm applied to 50
randomly generated datasets.  The $A$ matrices in state space models are chosen
to be random stable sparse matrices. Data is sampled from the simulation of
stochastic differential equations, with $1/h$ set to be close to but larger than
the critical sampling frequency by
Theorem~\ref{thm:identifi-largest-sampl-period}. The initial values of states
were randomly sampled from Gaussian distributions with zero mean, and the
process noise is Gaussian i.i.d, with SNR = 0 dB. Here, slightly abusing the
name of ``signals'', this ``SNR'' value is defined as
$\text{SNR} = 10 \log (\sigma_\text{init}^2/\sigma_\text{noise}^2)$, where
$\sigma_\text{init}^2$ denotes the variance of random initial states and
$\sigma_\text{noise}^2$ the variance of noise. Strictly speaking,
$\text{SNR} = -\infty\, \text{dB}$ since the initial state is unknown in
identification, and thus may not be treated as signals.  We choose low sampling
frequencies, large noise and limited samples to generate time series
challenging in identification, however, which is a typical profile of time
series in biological applications (e.g. microarray data \cite{He2012}).

The random generation of sparse stable $A$ matrices is not a trivial task. Due
to the lack of standards, it deserves time to explain our strategy to generate
random $A$ matrices. First, we do not want the network to be separable (i.e. a
collection of separate small networks). Thus, we first generate a loop of 24
nodes, which is represented by a stable $A$ matrix with nonzero diagonal and
up-right (or bottom-left) corner elements. It serves as a base to build up
$A$. Next a sparse matrix of the same dimension is generated with a fixed
sparsity density. The $A$ matrix is finally obtained by overlapping the sparse
matrix and the base and then permuting rows and columns randomly. During the
operation of overlapping two matrices, it might be possible that the combined
matrix is no longer stable. Therefore, we need a test of stability before
releasing $A$ matrices. If the matrix turns into unstable, we simply discard it
and search the next.

\begin{figure}[htbp]   %htbp!H
  \centering
  % \hspace*{-10mm}
  \includegraphics[width=.48\textwidth]{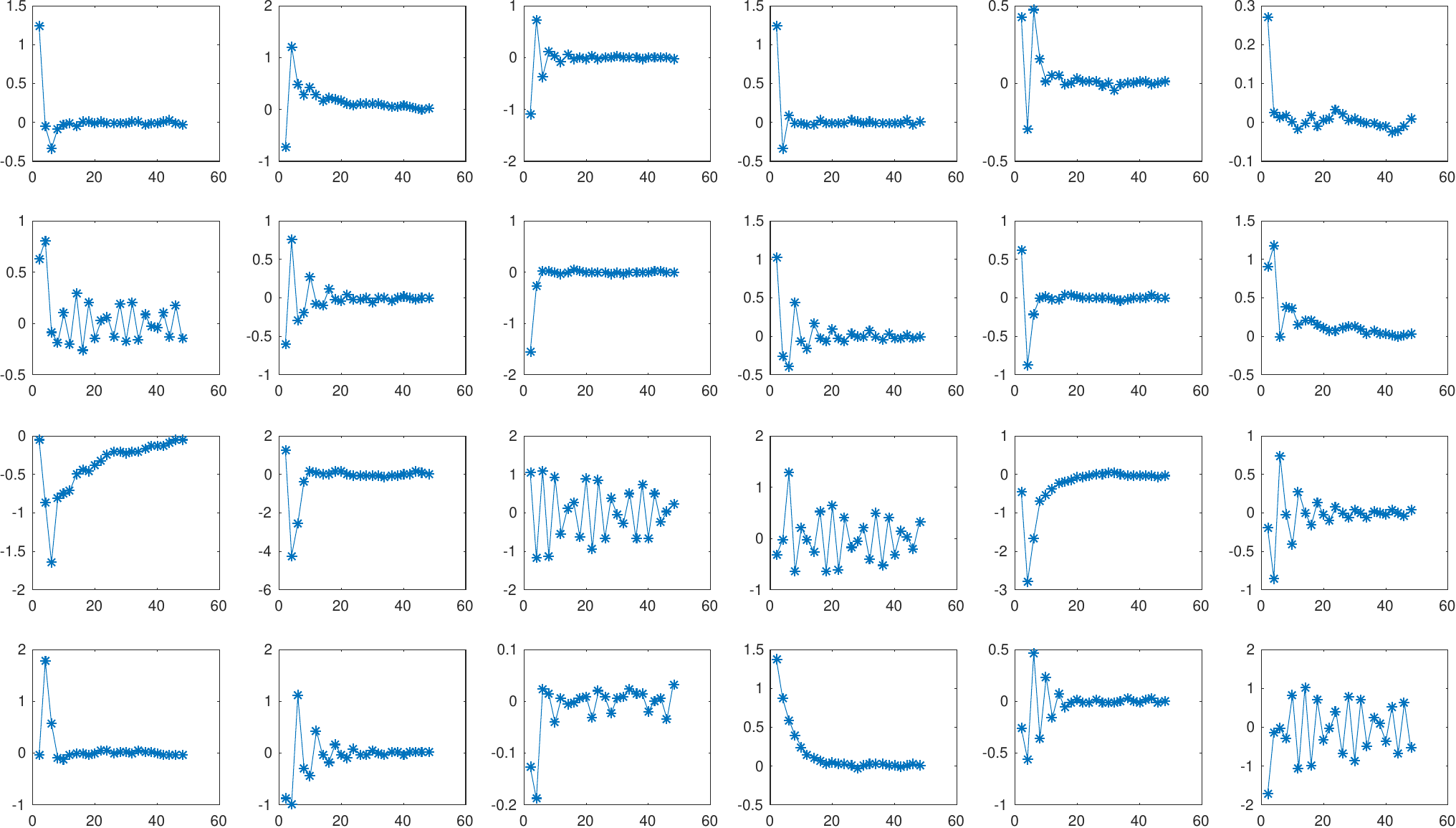}
  \caption{An example of time series used for reconstruction: 24 outputs, 24
    samples, random initial states, no inputs, and SNR = 0 dB.}
  \label{fig:output-signals}
\end{figure}

\begin{figure}[htbp]
  \centering
  \begin{subfigure}[b]{0.48\textwidth}
    \includegraphics[width=\textwidth]{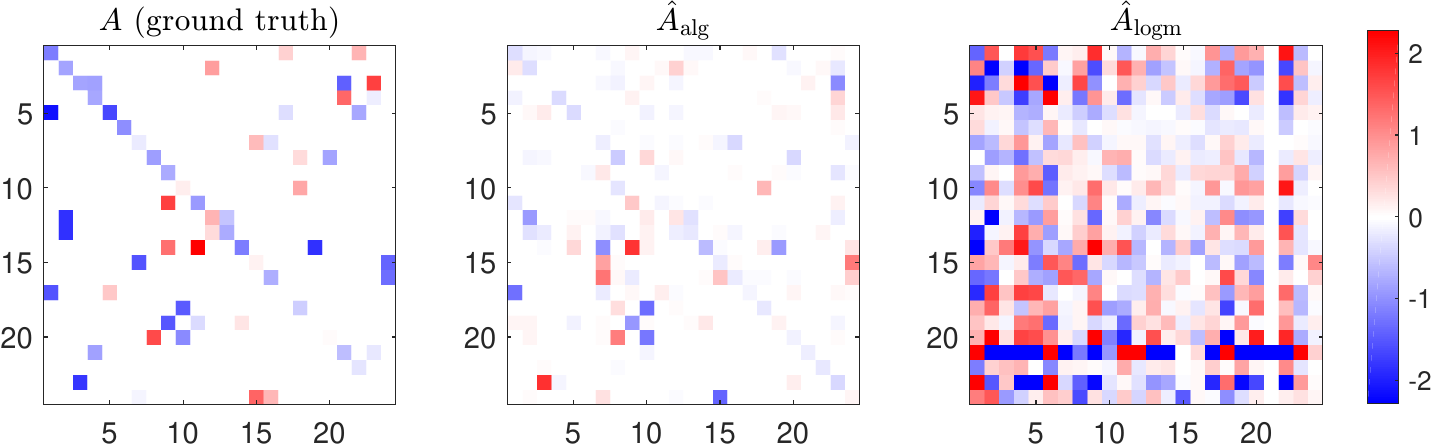}
    \caption{Ground truth and the estimations of $A$}
    \label{fig:subfig-A}
  \end{subfigure}
  \\ %add desired spacing between images, e. g. ~, \quad, \qquad, \hfill etc.
  \begin{subfigure}[b]{0.48\textwidth}
    \includegraphics[width=\textwidth]{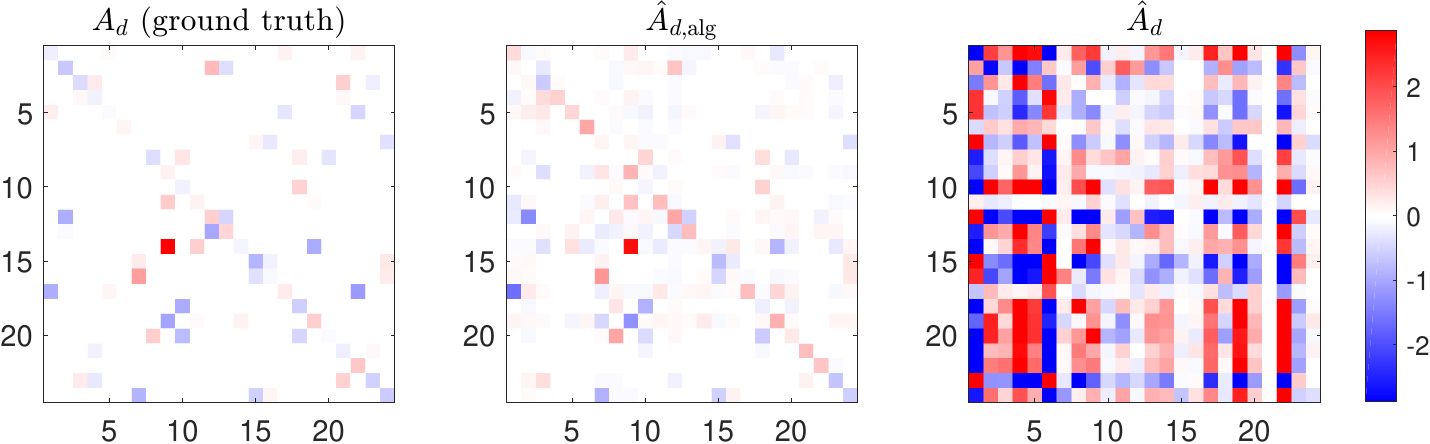}
    \caption{Corresponding $A_d$ calculated by $\exp(hA)$}
    \label{fig:subfig-Ad}
  \end{subfigure}
  \caption{An example of network reconstruction results.  $A$ and $A_d$ are the
    ground truth; $\hat{A}_\mathrm{alg}$ is estimated by the proposed method
    using $\lambda = 0.01$, and $\hat{A}_{d, \mathrm{alg}}$ is calculated by
    $\exp(h \hat{A}_{\mathrm{alg}})$; $\hat{A}_\mathrm{logm}$ is the principal
    logarithm of $\hat{A}_{d}$ divided by $h$, and $\hat{A}_d$ is estimated by
    PEM or ML.  % $\hat{A}_{d} = X_+ X_-^T (X_- X_-^T)^{-1}$
  }
  \label{fig:matrix-plot-A-Ad}
\end{figure}

\begin{figure}[htb]   %htbp!H
  \centering
  \includegraphics[width=.42\textwidth]{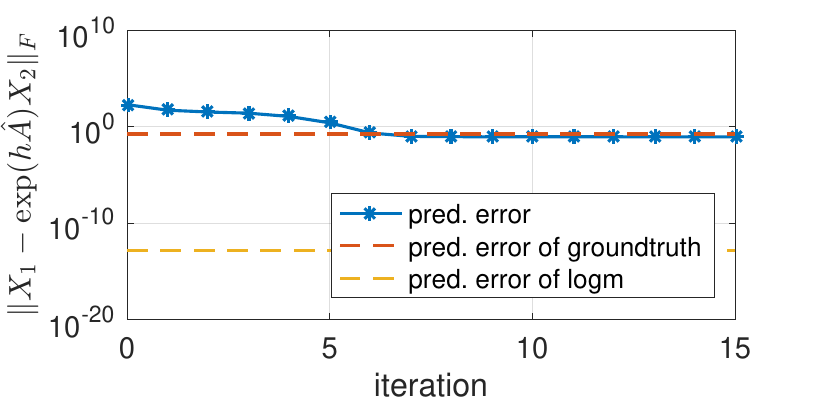}
  \caption{An example of convergence curve, with $\lambda = 0.01$ and zero as
    the initial point, for the chosen data set in
    Figure~\ref{fig:output-signals}.}
  \label{fig:convergence-alg}
\end{figure}

\begin{figure}[htbp]
  \centering
  \begin{subfigure}[b]{0.23\textwidth}
    \includegraphics[width=\textwidth]{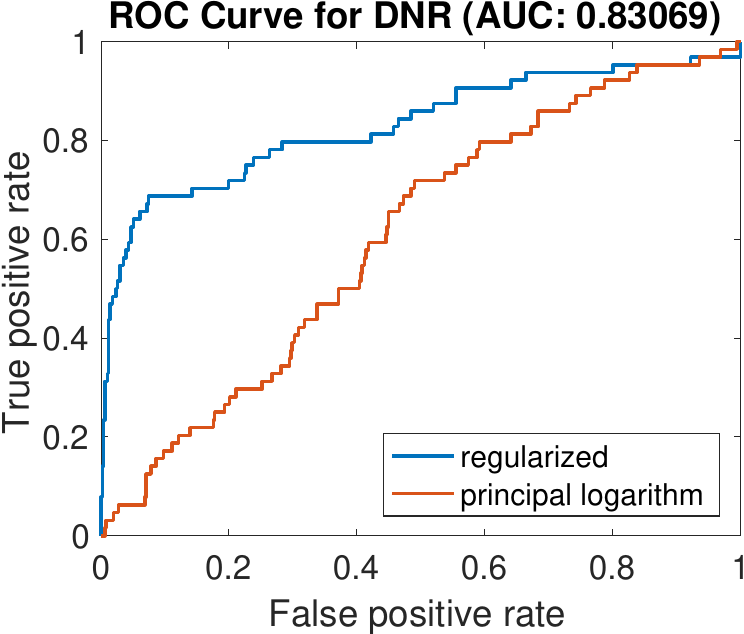}
    \caption{ROC curve}
    \label{subfig:ROC}
  \end{subfigure}
  ~
  \begin{subfigure}[b]{0.23\textwidth}
    \includegraphics[width=\textwidth]{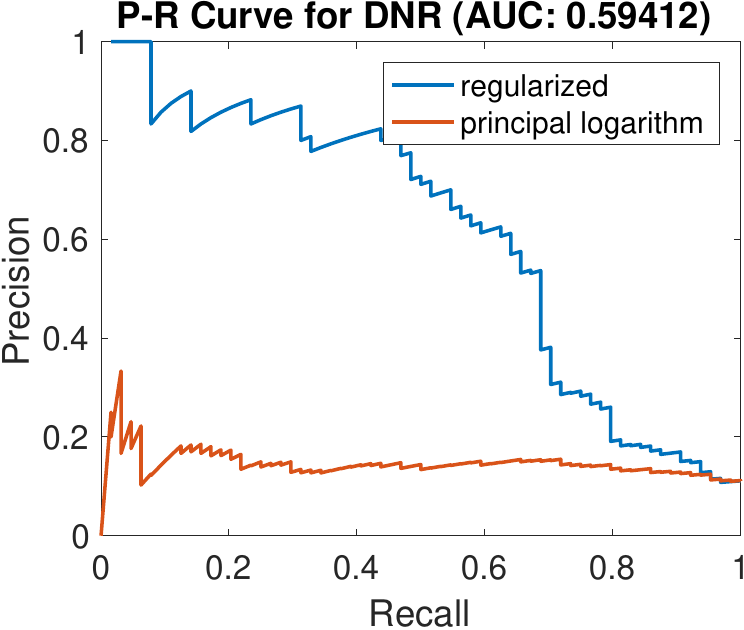}
    \caption{P-R curve}
    \label{subfig:PR}
  \end{subfigure}
  \caption{The ROC and the Precision-Recall curves for the chosen example.}
  \label{fig:ROC-PR-curve}
\end{figure}

An example of time series is given in Figure~\ref{fig:output-signals}.
The reconstruction results of this dataset is shown in
Figure~\ref{fig:matrix-plot-A-Ad}, together with the corresponding $A_d$'s
computed via matrix exponential. The straightforward way to estimate $A$ is
taking the principal matrix logarithm of PEM/ML solution $\hat{A}_{d}$,
which is, however, contaminated by process noise and unable to give reasonable
sparse structure of $A$, clearly shown as $\hat{A}_{\mathrm{logm}}$in
Figure~\ref{fig:subfig-A}. Taking matrix logarithm of least square estimations
of $A_d$ mostly encounters the issue of non-existence of principle logarithms,
which results in complex values of $A$. This shows the effects of process noise
on then estimation through matrix logarithms. However, the direct logarithm of
$\hat{A}_d$ might also work well when the dimension is small
(e.g. $\mathrm{dim}(A) \leq 6$).
The curve of prediction errors is shown in Figure~\ref{fig:convergence-alg},
which shows the convergence behaviors of
Algorithm~\ref{alg:iter-algor-optimisation}. Here the $\lambda$ is chosen by
performing network reconstruction on one dataset using $\lambda$ logarithmically
ranging from $10^{-4}$ to $100$ and checking the sparsity of $\hat{A}$'s (users'
prior knowledge) and the resultant prediction errors (whiteness, mean, standard
deviations). This value of $\lambda$ is then applied to all the other
datasets. Indeed, the choice of $\lambda$ also depends on $A$'s, which, however,
is randomly generated with the same sparsity. That might explain why the same
$\lambda$ works almost well for all datasets. Alternatively, $\lambda$ could be
automatically calculated by running the cross-validation technique, when the
amount of data allows.  As widely used in bioinformatics, the Receiver Operating
Characteristic (ROC) curve and the Precision-Recall (P-R) curve of this example
are provided in Figure~\ref{fig:ROC-PR-curve}.

\begin{figure}[htbp]
  \centering
  \begin{subfigure}[b]{0.24\textwidth}
    \includegraphics[width=\textwidth]{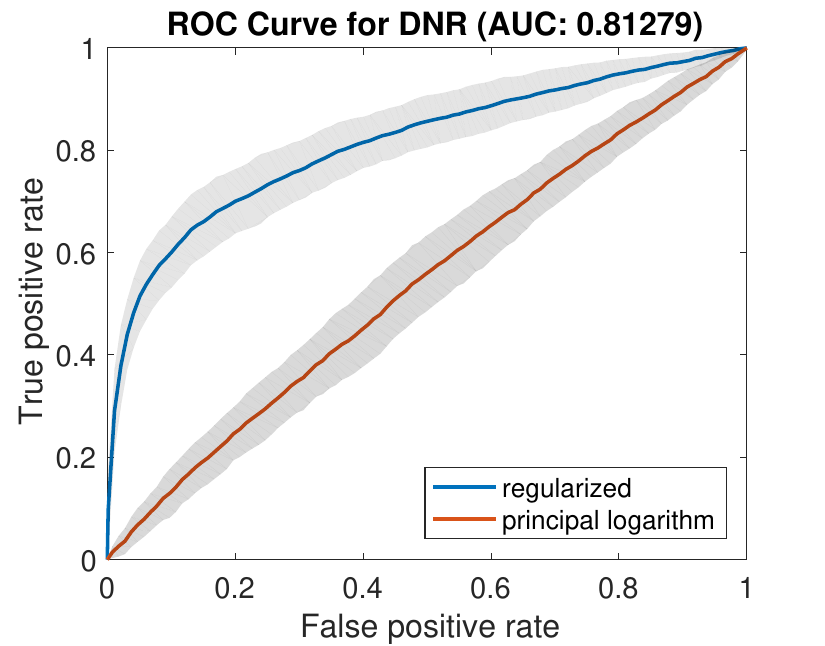}
    \caption{ROC curve}
    \label{subfig:ROC-avg}
  \end{subfigure}
  \begin{subfigure}[b]{0.24\textwidth}
    \includegraphics[width=\textwidth]{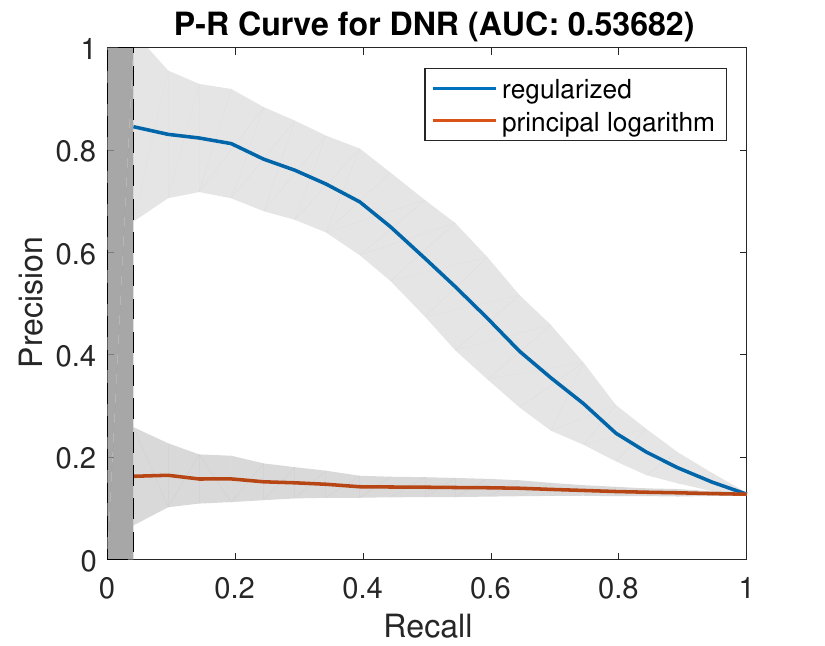}
    \caption{P-R curve}
    \label{subfig:PR-avg}
  \end{subfigure}
  \caption{The ROC and P-R curves averaging over the results of the proposed
    algorithm used on 50 random systems. The shaded area corresponds to one
    standard deviation from the mean value. The AUC values in titles refer to
    the mean of AUC values of the proposed method. The darkest area circled by
    dash lines is the uncertain region due to the average of \texttt{NaN} values
    in a few P-R curves.}
  \label{fig:ROC-PR-curve-avg}
\end{figure}

\begin{figure}[htbp]
  \centering
  \begin{subfigure}[b]{.24\textwidth}
    \includegraphics[width=\textwidth]{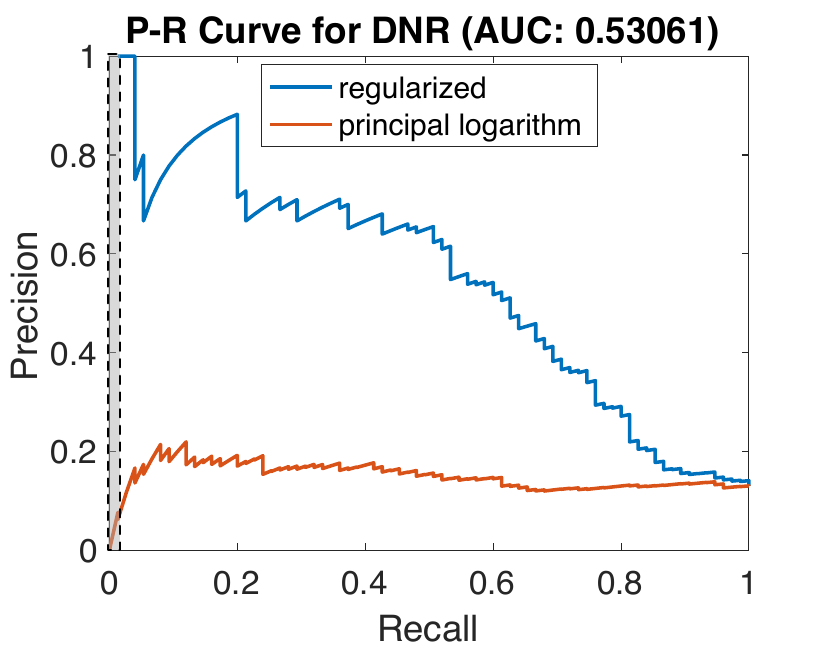}
    \caption{regular P-R curve}
    \label{subfig:ROC-reg}
  \end{subfigure}
  \begin{subfigure}[b]{.24\textwidth}
    \includegraphics[width=\textwidth]{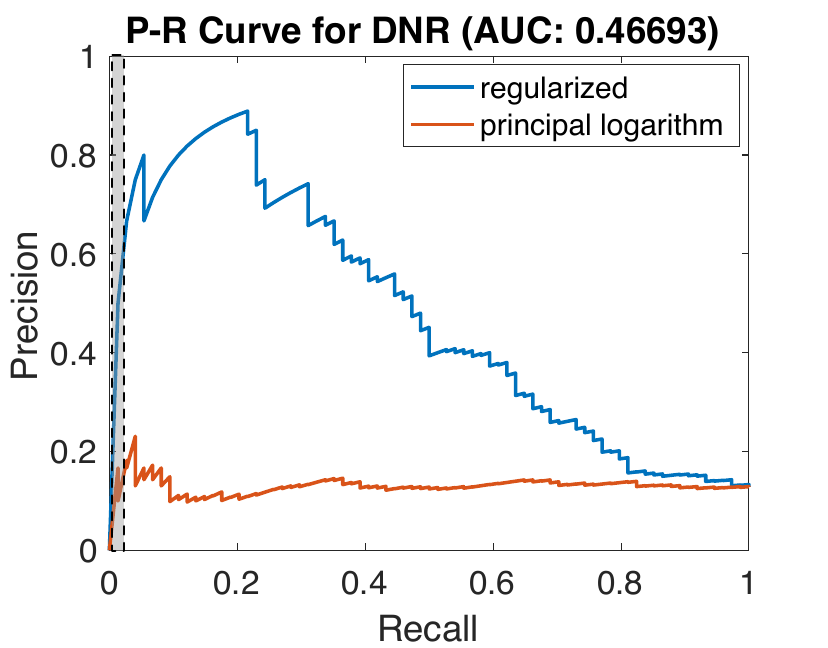}
    \caption{irregular P-R curve}
    \label{subfig:PR-irreg}
  \end{subfigure}
  \caption{Examples of P-R curves which show regular shapes and irregular
    shapes, which is supplement to the explanation of undefined region in
    Figure~\ref{subfig:PR-avg}. The shaded area in (a) is undefined region,
    which differs in different datasets; the one in (b) shows irregular
    profiles.}
  \label{fig:PR-PR-curve-reg}
\end{figure}

To show the performance of the proposed method, the ROC and the P-R curves,
averaging over reconstruction results of 50 random systems, are shown in
Figure~\ref{fig:ROC-PR-curve-avg}. The variables used in ROC and P-R curves are
computed by MATLAB function \texttt{perfcurve} with \texttt{XVals}
fixed. However, one has to notice that, at certain values of ``Recall'' close to
0, the corresponding ``Precision'' is not defined, as shown in
Figure~\ref{subfig:ROC-reg}, due to the fixed \texttt{XVals}. However, we need
to fixed the value of \texttt{XVals} in order to take average of 50 P-R
curves. One may notice the irregular profile of P-R curves for certain datasets,
where the corresponding values of ``Precision'' drop to zero in the
neighborhood of zero ``Recall''.

The numerical examples are programmed and computed in MATLAB, and the codes will
be released in public on the
\href{https://github.com/oracleyue}{github.com/oracleyue}. Considering
computational efficiency, we directly used vector/matrix norms instead of
quadratic forms in implementation (cf. \cite[chap.~11.1]{cvx-manual}).  Thanks
to the \emph{matrix function toolbox} \cite{matrix-function-toolbox} and the
\emph{CVX} \cite{cvx} for easy usage of matrix functions and convex
optimisation in MATLAB.

\section{Conclusions}
\label{sec:conclusions}

Continuous-time system identification is challenging when only
low-sampling-frequency data with limited lengths are available. Unfortunately,
this is a typical profile of time series in biomedicine applications. To
reconstruct the correct dynamic networks from such time series, we have to
identify the continuous-time models with sparse network structures.  This paper
studies the full-state measurement case, which is supposed to be the basic case
while it shows particular complications. We first clarify the concept of system
aliasing, which is raised by low sampling frequencies. A theorem on how to
choose the sampling frequency to guarantee no system aliasing is provided,
together with a test criterion. In regard to the ``easy'' case, i.e. no system
aliasing, we present an algorithm to reconstruct sparse dynamic network from
full-state measurements. In the case with system aliasing, the possibility on
searching among system aliases is manifested in theory relying on the prior
information on network sparsity. The paper dedicates to show the challenges and
attract more people contributing to this study.

% --------------- APPENDIX --------------
\appendices

\section{Matrix Exponential and Logarithm}

\begin{theorem}[Gantmacher{\protect\cite[Thm.~1.27]{Higham2008}}]
  \label{thm:matr-logar-Gantmacher}
  Let $P \in \mathbb{C}^{n \times n}$ be nonsingular with the Jordan canonical form
  \begin{subequations}
    \label{eq:jordan-form-matrix}
    \begin{alignat}{3}
      Z^{-1} P Z & = J = \diag(J_1, J_2, ..., J_p)\\
      J_k & = J_k(\lambda_k) =
      \begin{bmatrix}
        \lambda_k & 1 & & \\
        & \lambda_k & \ddots & \\
        & & \ddots & 1 \\
        & & & \lambda_k
      \end{bmatrix} \in \mathbb{C}^{m_k \times m_k}.
    \end{alignat}
  \end{subequations}
  Then all solutions to $e^{A} = P$ are given by
  \begin{equation}
    \label{eq:mat-log-all-sols-main}
    A = Z U \diag(L_1^{j_1}, L_2^{j_2}, ..., L_p^{j_p}) U^{-1} Z^{-1},
  \end{equation}
  where
  \begin{equation}
    \label{eq:mat-log-all-sols}
    L_k^{j_k} = \log(J_k(\lambda_k)) + 2 j_k \pi i I_{m_k};
  \end{equation}
  $\log(J_k(\lambda_k))$ denotes
  \begin{displaymath}
    f(J_k) \coloneqq
    \begin{bmatrix}
      f(\lambda_k) & f'(\lambda_k) & \cdots & \frac{f^{(m_k-1)}(\lambda_k)}{(m_k - 1)!} \\
      & f(\lambda_k) & \ddots & \vdots \\
      & & \ddots & f'(\lambda_k) \\
      & & & f(\lambda_k)
    \end{bmatrix}
  \end{displaymath}
  with $f$ the principal branch of the logarithm, defined by $\mathrm{Im}(\log(z)) \in (-\pi, \pi]$; $j_k$ is an arbitrary integer; and $U$ is an arbitrary nonsingular matrix that commutes with $J$.
\end{theorem}

\begin{theorem}[classification of logarithms{\protect\cite[Thm.~1.28]{Higham2008}}]
  \label{thm:matrix-logarithm-classification}
  Let the nonsingular matrix $P \in \mathbb{C}^{n \times n}$ have the Jordan canonical form~\eqref{eq:jordan-form-matrix} with $p$ Jordan blocks, and let $s \leq p$ be the number of distinct eigenvalues of $A$. Then $e^A = P$ has a countable infinity of solutions that are primary functions of $P$, given by
  \begin{equation}
    \label{eq:logm-classifiction-primary}
    A_j = Z \diag(L_1^{j_1}, L_2^{(j_2)}, ..., L_p^{(j_p)}) Z^{-1},
  \end{equation}
  where $L_k^{j_k}$ is defined in \eqref{eq:mat-log-all-sols}, corresponding to all possible choices of the integers $j_1, ..., j_p$, subject to the constraint that $j_i = j_k$ whenever $\lambda_i = \lambda_k$.

  If $s < p$ then $e^A = P$ has nonprimary solutions. They form parametrized families
  \begin{equation}
    \label{eq:logm-classification-nonprimary}
    A_j(U) = ZU \diag(L_1^{j_1}, L_2^{(j_2)}, ..., L_p^{(j_p)}) U^{-1}Z^{-1},
  \end{equation}
  where $j_k$ is an arbitrary integer, $U$ is an arbitrary nonsingular matrix that commutes with $J$, and for each $j$ there exist $i$ and $k$, depending on $j$, such that $\lambda_i = \lambda_k$ while $j_i \neq j_k$.
\end{theorem}

\begin{definition}[Fr\'{e}chet Derivatives\cite{Higham2008}]
  \label{def:Frechet-devrivative}
  The Fr\'{e}chet derivative of the matrix function \(f: \mathbb{C}^{n\times n} \rightarrow \mathbb{C}^{n \times n} \) at a point \(X\in \mathbb{C}^{(n \times n)}\) is a linear mapping
  \begin{align*}
    \mathbb{C}^{n \times n} &\overset{L}{\longrightarrow} \mathbb{C}^{n \times n}\\
    E &\longmapsto L(A, E)
  \end{align*}
such that for all $E \in \mathbb{C}^{n\times n}$
  \[ f(A+E) - f(A) - L(A, E) = o(\| E \|). \]
\end{definition}

\medskip
The Fr\'echet derivative is unique if it exists, and for matrix functions $\mathrm{exp}$ (matrix exponential) and $\mathrm{Log}$ (principal matrix logarithm) it exists.
The Fr\'echet derivative of the function $\mathrm{exp}$ \cite{Higham2008} is
\begin{equation}
L_{\mathrm{exp}}(X,E) = \int_0^1e^{X(1-s)}Ee^{Xs}ds,
\end{equation}
which can be efficiently calculated by the \emph{Scaling-Pade-Squaring} method in \cite{Al-Mohy2009}.
It gives a linear approximation of $\exp$ at a given point $A_c$ in the direction $E$
\begin{equation}
  \label{eq:exp-At-Frechet-deriv}
  e^{h A} = e^{h({A}_c + E)} = e^{h{A}_c} + L(h{A}_c, hE) + O(\|hE\|^2).
\end{equation}
The Fr\'echet derivative of the function $\mathrm{Log}$ \cite{Higham2008} is
\begin{equation*}
L_{\mathrm{Log}}(X,E) = \int_0^1(t(X-I) + I)^{-1}E(t(X-I)+I)^{-1})dt,
\end{equation*}
and its efficient computation algorithm is provided in \cite{Al-Mohy2013}.

\begin{theorem}[Kronecker representation {\protect \cite[Thm.~10.13]{Higham2008}}]
  \label{thm:kron-repr-frechet-deriv}
  For $A \in \mathbb{C}^{n\times n}$, $\kvec(L(A,E)) = K(A) \kvec(E)$, where $K(A) \in \mathbb{C}^{n^2 \times n^2}$ has the representations
  \begin{subequations}
    \label{eq:vec-Frechet-deriv-exp}
    \begin{empheq}[left={K(A)=\empheqlbrace\,}]{align}
      \nonumber
      & (I \otimes e^A) \psi\left(A^T \oplus (-A) \right) \\
      \nonumber
      & (e^{A^T/2} \otimes e^{A/2}) \sinch\left(\frac{1}{2}[A^T \oplus (-A)]\right) \\
      \nonumber
      & \frac{1}{2}(e^{A^T} \oplus e^A) \tau\left(\frac{1}{2}[e^T \oplus (-A)]\right)
    \end{empheq}
  \end{subequations}
  where $\psi(x) = (e^x-1)/x$ and $\tau(x) =\tanh(x)/x$. The third expression is valid if \( \frac{1}{2} \| A^T \oplus (-A) \| < \pi /2 \) for some consistent matrix norm.
\end{theorem}
Here the operator $\oplus$ is the \emph{Kronecker sum}, defined as \[ (A \oplus B) = A\otimes I_n + I_m \otimes B, \] for $A \in \mathbb{R}^{m \times m}$ and $B \in \mathbb{R}^{n \times n}$.

\section{Test criteria on system aliasing}
\label{appdix:test-criteria-sys-alias}

\subsection{A test criterion for the cases with inputs}
\label{appdix:subsec:test-criteria-case-input}

\begin{proposition}
  \label{prop:pred-error-diff-distribution-B}
  Consider the dynamical system \eqref{eq:dyna-sys-ss-cont}, and two sampling periods
  $h_1, h_2 \in \mathbb{R}_+$ such that $h_2/h_1 \notin \mathbb{N}$. Let
  $f(A,B,h,t_k) \triangleq x(t_k) - \exp(hA) x(t_{k-1}) - \int_0^h \exp(sA)B
  \,\mathrm{d}s\, u(t_{k-1})$, $\hat{A} = \Log(\exp(h_1 A))/h_1$ and $\hat{B}$ be a value
  such that $\mathbb{E}(f(\hat{A}, \hat{B}, h_1, t_k) | x(t_{k-1}), u(t_{k-1})) = 0$
  for all $t_k$. The one-step prediction errors w.r.t. $h_2$ are defined as
  $\epsilon(t_k) = f(A,B,h_2,t_k)$,
  $\hat{\epsilon}(t_k) = f(\hat{A}, \hat{B}, h_1, t_k)$.  If
  \begin{equation}
    \label{eq:pred-err-diff-0-B-condition}
    \begin{array}{l@{}l}
      \mathbb{E}\Big(
      &\big( \exp(h_2 A) - \exp(h_2 \hat{A}) \big) x(t_k)\ + \\
      &\displaystyle\int_0^{h_2} \big( \exp(sA)B -
      \exp(s\hat{A})\hat{B} \big) \mathrm{d}s\, u(t_k) \Big) \neq 0,
    \end{array}
  \end{equation}
  we have
  \begin{equation*}
    \mathbb{E}(\epsilon(t_k)) = 0, \quad
    \mathbb{E}(\hat{\epsilon}(t_k)) \neq 0.
  \end{equation*}
\end{proposition}
The proof follows trivially by evaluating the expectation
$\mathbb{E}(\hat{\epsilon}(t_k))$. We no longer require $\mathbb{E}(x(t_k)) \neq 0$
since we can take advantages of $u(t_k)$ to satisfy
\eqref{eq:pred-err-diff-0-B-condition}. In the cases with $\mathbb{E}(x(t_k)) = 0$
and non-zero inputs in expectation $\mathbb{E}(u(t_k)) \neq 0$, assuming $A, \hat{A}$
are non-singular, the condition \eqref{eq:pred-err-diff-0-B-condition} can be further
simplified as
\begin{equation}
  \label{eq:pred-err-diff-0-B-condition-special}
  \begin{array}{@{}l@{}l}
    &\big( \exp(h_2 \hat{A}) -I \big)
      \big( \exp(h_1 \hat{A}) -I \big)^{-1}
      \big( \exp(h_1 {A}) -I \big) \\
    &\neq \exp(h_2 A) - I.
  \end{array}
\end{equation}
The simplification follows from \eqref{eq:pred-err-diff-0-B-condition} by noticing
$\int_0^h e^{x s} \mathrm{d}s = (e^{h x} - 1)/x$ and $A$ commutes with its matrix functions.

\subsection{Proof Lemma~\ref{lemma:diff-A-h1h2}}
\label{appdix:subsec:proof-lemma-A-h1h2}

\begin{proof}
  Let $A_d \coloneqq \exp(h_1 A)$, which has the Jordan canonical form
  \eqref{eq:jordan-form-matrix} (i.e. let $P$ in
  Theorem~\ref{thm:matr-logar-Gantmacher}) be $A_d$). By
  Theorem~\ref{thm:matr-logar-Gantmacher} and
  \ref{thm:matrix-logarithm-classification}, we have
  \begin{displaymath}
    \begin{array}{l@{\,}l@{\,}l@{}l}
      h_1A &= Z \diag(L_1^{j_1}, &\dots, L_p^{j_p}&) Z^{-1}, \\
      h_1\hat{A} &= Z \diag(L_1^{0}, &\dots, L_p^{0}&) Z^{-1}.
    \end{array}
  \end{displaymath}
  To compare $\exp(h_2 \hat{A})$ with $\exp(h_2 A)$, we need to find
  their Jordan canonical form by the definition of matrix exponential. To calculate the
  eigenvalues of $h_2\hat{A}$, consider the determinant
  $|\hat{\mu} I - h_2 \hat{A} | = 0 \Leftrightarrow |\hat{\mu}' I - h_1
  \hat{A} | = 0$, where $\hat{\mu}' \triangleq h_1/h_2 \hat{\mu}$ and
  $|\hat{\mu}' I - h_1 \hat{A} | = |\hat{\mu}' I - \diag(L_1^0, \dots,
  L_p^0)|$. It is equivalent to solve $p$ equations
  $|\hat{\mu}_k' I_k - L_k^0| = 0\; (k = 1,\dots,p)$, where $I_k$ denotes the
  identity matrix of the dimension compatible with $L_k^0$. It yields that
  \begin{displaymath}
    \begin{array}{l@{}l}
      &|\hat{\mu}_k' I_k - L_k^0| = |\hat{\mu}_k' I_k - \log(J_k(\lambda_k))| \\
      &= \Big| \hat{\mu}_k' I_k -
        \begin{bmatrix}
          \log(\lambda_k) & \log'(\lambda_k) & \cdots & * \\
          & \log(\lambda_k) & \ddots & \vdots \\
          & & \ddots & \log'(\lambda_k)\\
          & & & \log(\lambda_k)
        \end{bmatrix}
                \Big| \\
      &= \left( \hat{\mu}_k' - \log(\lambda_k) \right)^{m_k} = 0,
    \end{array}
  \end{displaymath}
  where $J_k, \lambda_k, j_k, m_k, I_{m_k}$ are given in
  \eqref{eq:jordan-form-matrix}, \eqref{eq:mat-log-all-sols}; and hence
  $\mu_k = h_2/h_1 \log(\lambda_k)$ with geometric multiplicity $m_k$. Similarly for
  $h_2A$, consider
  $|\mu_k' I_k - L_k^{j_k}| = |\mu_k' I_k - \log(J_k(\lambda_k)) - 2 j_k \pi i
  I_{m_k}| = 0$, where $\mu_k' \triangleq h_1/h_2 \mu_k$, $\mu$ is the eigenvalues of
  $h_2A$, the integer $j_k$ is given in \eqref{eq:mat-log-all-sols}, and $i$ is the
  imaginary unit. It yields $m_k = h_2/h_1 (\log(\lambda_k) + 2j_k \pi i )$ with
  multiplicity $m_k$. Considering the special forms of $L_k^{j_k}$, we
  have the following Jordan decomposition
  \begin{displaymath}
    \begin{array}{l@{\,}l}
      \diag(L_1^0, \dots, L_p^0) &= U \diag(J_1(\hat{\mu}_1), \dots,
                                   J_p(\hat{\mu}_p)) U^{-1}, \\[1mm]
      \diag(L_1^{j_1}, \dots, L_p^{j_p}) &= U \diag(J_1({\mu}_1), \dots,
                                           J_p({\mu}_p)) U^{-1},
    \end{array}
  \end{displaymath}
  where $J_k(\hat{\mu}_k)$ and $J_k(\mu_k)$ denote the corresponding Jordan blocks.
  Therefore, $\exp(h_2 A) = \exp(h_2 \hat{A})$ is equivalent to $\exp(J_k(\mu_k)) =
  \exp(J_k(\hat{\mu}_k))$ for any $k = 1,\dots,p$, which implies $\exp(h_2/h_1
  2j_k\pi i) = 0$. It leads to the conditions: $h_2/h_1 \in \mathbb{N}_+$, or $j_k
  \equiv 0, \forall k$ (i.e. $\hat{A} = A$).
\end{proof}

\subsection{Proof of Proposition~\ref{prop:pred-error-diff-distribution}}
\label{appdix:subsec:proof-prop-pred-err-diff-dist}

\begin{proof}
  Considering the dynamical system \eqref{eq:dyna-sys-ss-cont}, it is obvious that
  $\mathbb{E}(\epsilon(t_k)|x(t_{k-1})) = 0$. Now we evaluate the other expectation
  \begin{displaymath}
    \begin{array}{l@{\,}l}
      &\mathbb{E}\big(\hat{\epsilon}(t_k) \big) = \mathbb{E}\big(x(t_{k+1}) - \exp(h_2 \hat{A})x(t_k)\big)\\
      &= \mathbb{E}\Big(
        \begin{array}[t]{@{}l@{}l}
          &\big(x(t_{k+1}) - \exp(h_2 A)x(t_k) \big) +\\
          &\big( \exp(h_2A) - \exp(h_2 \hat{A}) \big) x(t_k) \Big)
        \end{array}\\
      &= 0 + (\exp(h_2A) - \exp(h_2 \hat{A})) \mathbb{E}(x(t_k)) \neq 0,
    \end{array}
  \end{displaymath}
  by Lemma~\ref{lemma:diff-A-h1h2}.
\end{proof}

\section{More details on the optimization problems}
\label{appdix:derivation-Pprim-P}

\subsection{Equivalent forms of $(P_1), (P_2)$}
\label{appdix:subsec:equivalent-forms-P1-P2}

The following shows how to derive $(P_1')$ from $(P_1)$. Consider $g
\in \partial f(A_k)$ in $(P_1)$, which implies $g = \nabla \phi(A_k) + \lambda z =
\nabla \phi(A_k) + \lambda \sgn(A_k) + \lambda(z - \sgn(A_k)) \triangleq \bar{g} +
\lambda(z-\sgn(A_k))$. Recall that $\max_{s \in [-1,1]^n} s^T x = \|x\|_1$ where $x$
is an $n$-dimensional vector. Then we have
\begin{equation*}
  \begin{array}{l@{\;}l}
    \displaystyle\sup_{g \in \partial f(A_k)} g^T p_k
    &= \bar{g}^T p_k + \lambda \displaystyle\sup_{z \in J_1 \times \cdots \times J_{n^2}} \left( z - \sgn(A_k) \right)^T p_k \\
    &= \bar{g}^T p_k + \lambda \displaystyle\sup_{\bar{z} \in [-1,1]^l} \bar{z}^T
      \left( W(A_k)p_k \right) \\
    &= \bar{g}^T p_k + \lambda \|W(A_k)p_k\|_1,
  \end{array}
\end{equation*}
where $W(A_k) = I - \diag(|\sgn(A_k)|)$ and $l = n^2 - \operatorname{card}(\sgn(A_k))$.

Consider the optimization $(P_2)$, we go through the same procedure and obtain
\begin{equation*}
  \sup_{g \in \partial f(\theta)} g^T p_k = \bar{g}^T p_k + \lambda \|W(A_k) \Lambda p_k\|_1,
\end{equation*}
where $\bar{g} \triangleq \nabla\phi(\theta) + \lambda \Lambda^T \sgn(A_k)$.

\subsection{Proof of Proposition~\ref{prop:no-zero-p-before-local-optima}}
\label{appdix:subsec:proofs-proposition-nonzero}

\begin{proof}
  Without loss of generality, suppose that there exists $p^*_k \neq 0$ and
  $0 \notin \operatorname{arg\,min}_{p_k} \hat{f}(A_k,p_k)$ such that
  $\sup_{g \in \partial f(A_k)} g^T p^*_k = 0$ and $0 \notin \partial f(A_k)$.
  (Indeed, if $0 \in \operatorname{arg\,min}_{p_k} \hat{f}(A_k,p_k)$, we apply
  Proposition~\ref{prop:zero-p-guarantee-local-optima} and obtain
  $0 \in \partial f(A_k)$.)
  It implies that
  $\sup_{g \in \partial f(A_k)} g^T \alpha p^*_k = 0, \:\forall \alpha \in (0,1]$.
  Hence, $f(A_k) \leq f(A_k+ \alpha p_k^*)$ for small enough $\alpha$.
  Let $\phi(A_k), \hat{\phi}(A_k, p_k)$ be defined by
  \begin{equation*}
    \begin{array}{r@{\;}l}
      f(A_k) &= \phi(A_k) + \lambda \|\kvec(A_k)\|_1, \\
      \hat{f}(A_k,p_k) &= \hat{\phi}(A_k,p_k) + \lambda
  \|\kvec(A_k)+p_k\|_1,
    \end{array}
  \end{equation*}
  and hence $\phi(A_k + \ikvec(p_k)) = \hat{\phi}(A_k, p_k) + o(\|p_k\|)$
  ($\|\cdot\|$ denotes any vector norm).  For simplicity, without any ambiguity,
  we use $f(A_k + p_k)$ to represent $f(A_k + \ikvec(p_k))$.  Then we have
  $f(A_k + p_k) = \hat{f}(A_k,p_k) + o(\|p_k\|)$, which yields
  \begin{equation}
    \label{eq:small-o-limit}
    \lim_{\alpha \downarrow 0} \frac{|f(A_k + \alpha p_k^*) - \hat{f}(A_k, \alpha
      p_k^*)|}{\alpha \|p_k^*\|} = 0.
  \end{equation}
  Now let us calculate this limit in a different way. Noting that $\hat{f}(A_k,
  p_k^*) \leq \hat{f}(A_k, \alpha p_k^*)$ (since $p_k^* \in
  \operatorname{arg\,min}_{p_k} \hat{f}(A_k,p_k)$) and
  $\hat{f}(A_k, p_k^*) < \hat{f}(A_k, 0) = f(A_k)$ (since $0 \notin
  \operatorname{arg\,min}_{p_k} \hat{f}(A_k,p_k)$), we have
  \begin{equation*}
    \hat{f}(A_k, \alpha p_k^*) - f(A_k) \geq \hat{f}(A_k, p_K^*) - f(A_k) \geq \delta > 0.
  \end{equation*}
  Moreover, since $f(A_k + \alpha p_k^*) - f(A_k) = \phi(A_k + \alpha
  p_k^*) - \phi(A_k) + \lambda (\|A_k + \alpha p_k^*\|_1 - \|A_k\|_1) = o(\alpha
  \|p_k^*\|) + O(\alpha \|p_k^*\|)$, there exists $M \in \mathbb{R}$ such that
  \begin{equation*}
    \lim_{\alpha \downarrow 0} \frac{\left|f(A_k + \alpha
        p_k^*) - f(A_k)\right|}{ \alpha \|p_k^*\|} \leq M.
  \end{equation*}
  Now we recalculate the limit in \eqref{eq:small-o-limit} as follows
  \begin{equation*}
    \begin{array}{l@{\;}l}
      &\displaystyle\lim_{\alpha \downarrow 0} \frac{|f(A_k + \alpha p_k^*) - \hat{f}(A_k, \alpha
      p_k^*)|}{\alpha \|p_k^*\|}\\
     &\geq \displaystyle\lim_{\alpha \downarrow 0} \left| \frac{\left|f(A_k + \alpha
        p_k^*) - f(A_k)\right|}{ \alpha \|p_k^*\|} - \frac{\left|f(A_k) -
        \hat{f}(A_k, \alpha p_k^*)\right|}{ \alpha \|p_k^*\|}   \right|\\
      &\geq \displaystyle \left| M - \lim_{\alpha \downarrow 0} \frac{\delta}{\alpha
             \|p_k^*\|}  \right|
      = +\infty,
    \end{array}
  \end{equation*}
  which contradicts with \eqref{eq:small-o-limit}.
\end{proof}

\subsection{Proof of Proposition~\ref{prop:zero-p-guarantee-local-optima}}
\label{appdix:subsec:proof-proposition-zero}

\begin{proof}
  Since $p_k^* \in \operatorname{argmin}_{p_k} \hat{f}(A_k,p_k)$, it yields
  $0 \in \partial_{p} \hat{f}(A_k, p_k^*)$. Note that, in our discussion, $A_k$ is
  always fixed, and thus $\partial_{p} \hat{f}(A_k,p_k)$ denotes the subgradient of
  $\hat{f}(A_k, \cdot)$ at $p_k$. Now let us write $\partial_{p} \hat{f}(A_k,p_k)$
  explicitly
  \begin{equation*}
    \partial_p \hat{f}(A_k,p_k) = \left\{ \nabla \hat{\phi}(A_k,p_k) + \lambda
      \hat{z}: \hat{z} \in \hat{J}_1 \times \cdots \hat{J}_{n^2} \right\},
  \end{equation*}
  where
  \begin{equation*}
    \hat{J}_i =
    \begin{cases}
      [-1,1] & \text{if } \kvec(A_k)_i + (p_k)_i = 0\\
      \{1\}  & \text{if } \kvec(A_k)_i + (p_k)_i > 0\\
      \{-1\} & \text{if } \kvec(A_k)_i + (p_k)_i < 0
    \end{cases},
  \end{equation*}
  $(p_k)_i$ denotes the $i$-th element of $p_k$,
  and
  $\nabla \hat{\phi}(A_k,p_k) = 2 J(A_k)^T \left( r(A_k) + J(A_k) p_k
  \right)$. Hence $\partial_p \hat{f}(A_k, 0) = \partial f(A_k)$.
  Therefore, $0 \in \partial_p\hat{f}(A_k,p_k^*)$ with $p_k^* = 0$ implies $0
  \in \partial f(A_k)$.
\end{proof}

\section{Proofs for boundness of system aliases}
\label{appdix:proof-lemmas}

\subsection{Proof of Lemma~\ref{lemma:logm-As-condition-equality}}
\label{appdix:proof-lemma-1}

\begin{proof}
  Let $A_{i} \coloneqq Z \diag(L_1^{j_1^{(i)}}, \cdots, L_p^{j_p^{(i)}}) Z^{-1}$
, where $i = 1,2$, $L_k^{j_k^{(i)}} \coloneqq \log(J_k(\lambda_k)) + 2 j_k^{(i)} \pi i I_{m_k}$, and all other notations are given in \eqref{eq:logm-classifiction-primary}.
  \begin{figure*}[htb]
      \begin{alignat}{1}
        \|h_Z({A}_{i})\|_F^2 - \|h_Z({A_0})\|_F^2 & = \ \trace\left(\diag^*(L_1^{j_1^{(i)}}, \cdots, L_p^{j_p^{(i)}})
           \diag(L_1^{j_1^{(i)}}, \cdots, L_p^{j_p^{(i)}}) \right) \nonumber\\
        &\quad\  - \trace\left(\diag^*(L_1^{(0)}, \cdots, L_p^{(0)})
           \diag(L_1^{(0)}, \cdots, L_p^{(0)}) \right)\nonumber\\
        &= \textstyle\sum_{k=1}^p \trace\left( L_k^{j_k^{(i)}*}L_k^{j_k^{(i)}} -
                                   L_k^{(0)*}L_k^{(0)} \right) \label{eq:Ai-Aj-Frobenius-norm} \\
        &= \textstyle\sum_{k=1}^p \trace\left( 2j_k^{(i)} \pi i \left(\log(J_k)^* - \log(J_k)\right) + 4\pi^2j_k^{(i)2} I_{m_k} \right) \nonumber\\
        &=\textstyle\sum_{k=1}^p 4\pi j_k^{(i)} m_k (\beta_k + j_k^{(i)})
        =4\pi\, \mathscr{I}(0, \textit{j}), \quad \textit{j} \triangleq [j_1^{(i)}, \dots, j_p^{(i)}]. \nonumber
      \end{alignat}
  \end{figure*}
  By using \eqref{eq:Ai-Aj-Frobenius-norm} for $A_1, A_2$, we obtain
  \begin{align*}
    &\|h_Z(A_1)\|_F = \|h_Z(A_2)\|_F \Leftrightarrow \|h_Z(A_1)\|_F^2 - \|h_Z(A_0)\|_F^2 \\
    & = \|h_Z(A_2)\|_F^2 - \|h_Z(A_0)\|_F^2 \Leftrightarrow \mathscr{I}(\textit{j}^{(1)}, \textit{j}^{(2)} - \textit{j}^{(1)}) = 0,
  \end{align*}
  which implies that $A_1 \sim A_2$ by definition. The first equality in \eqref{eq:Ai-Aj-Frobenius-norm} is due to the linear transformation $h_Z(\hat{A})$.
\end{proof}

\subsection{Proof of Lemma~\ref{lemma:finite-elements-equiv-class}}
\label{appdix:proof-lemma-2}

\begin{proof}
  Let $\textit{j}$ denote $[j_1, \dots, j_p]$ of $\bar{A}$ in \eqref{eq:mat-log-all-sols}, and $\textit{j}^{(i)}$ denotes $[j_1^{(i)},\dots,j_p^{(i)}]$ of $A_i \in {\mathcal{S}}$. $\delta \triangleq \textit{j}^{(i)} - \textit{j}$, therefore $\delta \in \mathbb{Z}^{p}$,
where $\succeq$ denote the element-wise larger-or-equal relation. By Definition~\ref{def:equivalence-relation}, it is equivalent to show that $\mathscr{I}(\textit{j}, \delta)=0$ has finite solutions, given $\textit{j}$. We require $\delta$ to satisfy the following condition:
\begin{equation}
  \label{eq:delta-bound-eqiv-class}
  |\delta_i + j_i + \beta_i/2| \leq \sqrt{\frac{(\textit{j}+\beta/2)^T M (\textit{j}+\beta/2)}{m_i}}
\end{equation}
for all $i = 1,\dots,p$.
Otherwise, supposing that there exists $i \in \{1,\dots, p\}$ such that $\delta_i$ does not satisfy \eqref{eq:delta-bound-eqiv-class}, we will have
\begin{align*}
  \mathscr{I}(\textit{j},\delta) &= m_i(\delta_i + j_i + \beta_i/2)^2 + \sum_{k \neq i} m_k(\delta_k + j_k \beta_k/2)^2 \\
                      &- \sum_k m_k (j_k + \beta_k/2)^2
                      > \sum_{k \neq i} m_k(\delta_k + j_k \beta_k/2)^2 \geq 0.
\end{align*}
Let ${\mathcal{S}}' \coloneqq \{A_i \in {\mathcal{S}}: j_k^{(i)} = \delta_k + j_k, \delta_k \text{ satisfies \eqref{eq:delta-bound-eqiv-class}} \}$.
We have $\{A_i \in {\mathcal{S}}: A_i \sim \bar{A}\} \subseteq {\mathcal{S}}'$ and ${\mathcal{S}}'$ is a finite set.
\end{proof}

\subsection{Proof of Lemma~\ref{lemma:finite-A-Fro-less-k}}
\label{appdix:proof-lemma-3}

\begin{proof}
  Let $\kappa_0 \triangleq \|h_Z(A_0)\|_F$. Then we need to show there exists a finite number of $A_i \in {\mathcal{S}}$ such that $\|h_Z(A_i)\|_F^2 - \|h_Z(A_0)\|_F^2 \leq \kappa^2 - \kappa_0^2$, which is equivalent to show that there exists a finite number of solutions $\delta \in \mathbb{Z}$ to $\mathscr{I}(0, \delta) \leq (\kappa^2 - \kappa_0^2)/4\pi$. $\delta$ must satisfy the following condition:
  \begin{equation}
    \label{eq:delta-bound-k-upperbound}
    | \delta_i + \beta_i/2| \leq \sqrt{\frac{(\beta/2)^TM(\beta/2) + (\kappa^2 - \kappa_0^2)}{m_i}}
  \end{equation}
  for all $i =  1, \dots, p$. Otherwise, by supposing that there exists $i \in {1,\dots,p}$ such that $\delta_i$ does not satisfy \eqref{eq:delta-bound-k-upperbound}leads to
  \begin{align*}
    \mathscr{I}(0,\delta) &= m_i(\delta_i + \beta_i/2)^2 + \\
    & \sum_{k \neq i} m_k(\delta_k + \beta_k/2)^2 - (\beta/2)^T M (\beta/2) \\
                > & \sum_{k \neq i} m_k(\delta_k + \beta_k/2)^2 + (\kappa^2 - \kappa_0^2) \geq \kappa^2 - \kappa_0^2.
  \end{align*}
  Note that the set of all $\delta \in \mathbb{Z}$ that satisfies \eqref{eq:delta-bound-k-upperbound} is finite, which finalizes the proof.
\end{proof}

\subsection{Proof of Proposition~\ref{thm:identifi-boundness-A}}
\label{appdix:subsec:proof-proposition-boundedness}

\begin{proof}
  Let $\textit{j}$ denote $[j_1, \dots, j_p]$ of $\bar{A}$ in \eqref{eq:mat-log-all-sols}, $N_\mathrm{eqiv}$ be the number of $A$'s that satisfy $A \sim \bar{A}$.
  Note that $|\|h_Z(A)\|_F^2 - \|h_Z(\bar{A})\|_F^2| = |(\|h_Z(A)\|_F^2 - \|h_Z(A_0)\|_F^2) - (\|h_Z(\bar{A})\|_F^2 - \|h_Z(A_0)\|_F^2)| = |\mathscr{I}(\textit{j}, \delta)|, \delta \in \mathbb{Z}$, which implies it is equivalent to show that $|\mathscr{I}(\textit{j}, \delta)|, \delta \in \mathbb{Z}$ has a non-zero lower bound if not considering the $\delta$'s that result in $\mathscr{I}(\textit{j}, \delta) =0$. We will prove it by contradiction. Assume this is not true, i.e. $\forall \epsilon >0$ there exists $\delta$ such that $0 < |\mathscr{I}(\textit{j}, \delta)| < \epsilon$. It implies that, arbitrarily given $\epsilon > 0$, there exists an infinite number of $\delta$ such that $\mathscr{I}(\textit{j}, \delta) < \epsilon$, which is impossible since $\mathscr{I}(0,\textit{j}+\delta) < \mathscr{I}(0, \textit{j}) + \epsilon$ (using the fact that $\mathscr{I}(\textit{j},\delta) = \mathscr{I}(0,\textit{j}+\delta) - \mathscr{I}(0, \textit{j})$) has a finite number of solutions provided by Lemma~\ref{lemma:finite-A-Fro-less-k}.
\end{proof}

% --------------------------------------------------
% supplement figures
% --------------------------------------------------

% --------------- BIBLIOGRAPHY --------------
\bibliographystyle{IEEEtran}
\bibliography{./ref/library,./ref/ref}

\end{document}

%% file: userdef-mathsym.tex
\newcommand{\vertiii}[1]{{\left\vert\kern-0.25ex\left\vert\kern-0.25ex\left\vert #1
    \right\vert\kern-0.25ex\right\vert\kern-0.25ex\right\vert}}  %matrix norm

\newcommand{\diag}{\operatorname{diag}}

\newcommand{\trace}{\operatorname{tr}}
\newcommand{\im}{\operatorname{Im}}

\newcommand{\kvec}{\operatorname{vec}}
\newcommand{\ikvec}{\operatorname{ivec}}
\newcommand{\sinch}{\operatorname{sinch}}
\newcommand{\Log}{\operatorname{Log}}
\newcommand{\eig}{\operatorname{eig}}
\newcommand{\sgn}{\operatorname{sgn}}

\newcommand{\minimize}[1]{\underset{#1}{\operatorname{minimize}}}  % or using \,
\newcommand{\st}{\operatorname{subject\ to}}